\documentclass{article}

\usepackage{amssymb}
\usepackage{amsthm}
\usepackage{amsmath}
\usepackage{authblk}
\usepackage{blkarray}
\usepackage{booktabs}
\usepackage{bm}
\usepackage{calc}
\usepackage{caption}
\usepackage{cases}
\usepackage{cite}
\usepackage{dcolumn}
\usepackage{empheq}
\usepackage[margin=1.25in]{geometry}
\usepackage{graphicx}
\usepackage{hyperref}
\usepackage[utf8]{inputenc}
\usepackage{lineno}
\usepackage{listings}
\usepackage{mathtools}
\usepackage{newpxtext,newpxmath}
\usepackage[super]{nth}
\usepackage{setspace}
\usepackage{subcaption}
\usepackage{tabularx}
\usepackage{tikz}
\usepackage{tikzsymbols}
\usepackage{verbatim}
\usepackage{wrapfig}
\usepackage[theorems,skins]{tcolorbox}
\usepackage{csquotes}
    \MakeOuterQuote{"}

\newtcolorbox[auto counter]{mymathbox}[2][]{colback=white, sharp corners, ams align, colframe=blue!30!black, title=Box~\thetcbcounter: #1, label=#2}

\title{Suppressing evolution through environmental switching}
\date{\today}
\author[1,2]{Bryce Morsky \thanks{morsky@sas.upenn.edu}}
\author[1]{Dervis Can Vural \thanks{dvural@nd.edu}}
\affil[1]{Department of Physics, University of Notre Dame, Nieuwland Science Hall, Notre Dame, IN, USA}
\affil[2]{Department of Biology, University of Pennsylvania, Carolyn Lynch Laboratories, Philadelphia, PA, USA }

\newcommand\m[1]{\langle #1 \rangle}

\begin{document}

\maketitle

\begin{abstract}
Ecology and evolution under changing environments are important in many subfields of biology with implications for medicine. Here, we explore an example: the consequences of fluctuating environments on the emergence of antibiotic resistance, which is an immense and growing problem. Typically, high doses of antibiotics are employed to eliminate the infection quickly and minimize the time under which resistance may emerge. However, this strategy may not be optimal. Since competition can reduce fitness and resistance typically has a reproductive cost, resistant mutants' fitness can depend on their environment. Here we show conditions under which environmental varying fitness can be exploited to prevent the emergence of resistance. We develop a stochastic Lotka-Volterra model of a microbial system with competing phenotypes: a wild strain susceptible to the antibiotic, and a mutant strain that is resistant. We investigate the impact of various pulsed applications of antibiotics on population suppression. Leveraging competition, we show how a strategy of environmental switching can suppress the infection while avoiding resistant mutants. We discuss limitations of the procedure depending on the microbe and pharmacodynamics and methods to ameliorate them.
\end{abstract}


\section{Introduction}

Populations will face a variety of environmental fluctuations of both biotic and abiotic nature. Since phenotypes typically have different reproductive success in differing environments, the dynamics of these fluctuations can be crucial in determining phenotypic composition. Here, we consider the effects of varying environments on the emergence and maintenance of antibiotic resistance.

The rise of microbial resistance is a looming catastrophe, and prudential use of antimicrobials is a fundamental means to prevent it \cite{laxminarayan13}. Such strategies to limit the chance of resistance can be made at all levels of disease dynamics, from population level protocols to individual patient therapies. Studies of antibiotic resistance in vivo, in hospitals, and in the community at large using mathematical models can help address the pharmacodynamics, pharmacokinetics, and epidemiology of resistance \cite{lipsitch97, bonhoeffer97, austin99, czock07, gloede09, greulich17, nielsen13}. Such models have found use in effectively modelling real-world experimental data \cite{tam07, schmidt09, bhagunde11, nielsen11}. In particular, modelling has been used in identifying dosing regimens that suppress the emergence of resistance \cite{tam05, tam08}.

There are several mechanisms by which bacteria can be resistant to antibiotics \cite{poole02}, an example of which is overexpression of the efflux pump \cite{borges03, webber03, sun14}, which bacteria use to expel antibiotics. Typically, such resistant mechanisms have a fitness cost, which can result in trade-offs between resistance and growth \cite{martinez02,ender04,wang17,basra18}. Example costs of resistance include less energy for other cellular processes, and impaired motility. When the antibiotic is present, the resistant mechanism pays for its cost by providing a fitness advantage relative to the susceptible strain. However, when the antibiotic is not present, the resistant mechanism incurs a fitness disadvantage. Resistance, therefore, can be reversed under evolutionary forces by altering the environment \cite{andersson10}. As such, a pulsed protocol, where the antibiotic is periodically applied so that the environment switches from antibiotic to antibiotic-free regimes, may be able to eliminate the bacteria. However, there is a risk that resistant mutants evolve to reduce the fitness cost of resistance rather than lose the resistance mechanism, whereby they could be competitive whether the antibiotic is present or not \cite{pacheco17}. Further, mutations in regulatory genes can produce phenotypes of irreversible resistance \cite{vanBambeke00}. These risks can cripple pulsed protocols aimed at controlling the infection while preventing resistance. Preventing a sustained presence of resistance is therefore a high priority.

Here, we develop a mathematical model of pulsed protocols of antibiotic and antibiotic-free regimes, switching rapidly from one environment to the other, to control a bacterial population comparable to \textit{Escherichia coli}. We consider concentration-independent (i.e.\ time-dependent) bactericides such as $\beta$-lactams (e.g.\ penicillins and cephalosporins), which require high maintained concentrations to be effective \cite{shojaee00}. We assume that there is a maximum benefit to the concentration amount (due either to the pharmacodynamics of the bactericidal mechanism or tolerance of the patient to the antimicrobial). Therefore, we fix the dose concentration when applied, and find the proper periods for each regime that prevent the emergence of resistance while eliminating the infection.

Previous theoretical studies have shown that pulsed protocols of antibiotics can eliminate bacteria \cite{kussell05,cogan06,cogan12,acar19}. However, these studies feature only a ``persistent phenotype'', that neither grows nor dies under application of the antimicrobial agent \cite{balaban04, zhang12}. Bacteria may transition between the persister type and wild type, depending on the environmental conditions. The number of persisters remain at low levels and act as a staging ground for the bacteria to repopulate after the antimicrobial is removed. Pulsed protocols of antibiotics, however, can disrupt this process and lead to the elimination of the bacteria. Experimental studies have shown that pulsed protocols can be effective in controlling such a system \cite{sharma15}.

Although pulsed protocols can eliminate non-persister and persister colonies, they have more difficulty in eliminating colonies with resistant phenotypes that can grow when antibiotics are present. However, these protocols have been shown to be effective in containing an infection both theoretically and experimentally \cite{baker18, hansen20}. In such cases, antimicrobials can act as ecological disturbances and can be approximately as effective as a constant application of the antimicrobial in controlling the bacterial load while also diminishing the probability of the emergence of resistance \cite{baker18}. With short durations of high concentrations of drugs, the period under which resistance is selected for can be minimized.

The above studies have explored pulsed protocols in different ways: controlling persisters, and controlling emergence of resistance. The resistant strain we consider here does grow in the presence of antibiotics, and thus are not persisters. Our scenario is thus more similar to, and an extension of, \cite{baker18}. Our main contribution is to show how leveraging competition can not only suppress the emergence of resistance as in \cite{baker18}, but also reduce the overall bacterial load. Additionally, we explore the impact of other important mechanisms on pulsed protocols including the evolvability of the bacteria and the lethality of the antibiotic. We compare these results to a protocol of constant application. Though pulsed protocols can, on average, outperform a constant application, constant applications are more likely to completely eliminate the bacteria. However, they are also more likely to result in an uncontrolled population of resistant mutants. Thus, with pulsed protocols we aim to mitigate the emergence of resistance and reduce the risk of the evolutionary escape from the antibiotic.

The key mechanism of our models in suppressing the population is competition between the two phenotypes. Two common models of microbial competition are resource-competition models \cite{baker18} and the competitive Lotka-Volterra equation \cite{stein13,gonze18}. The latter of which we employ here. Competition can be low when the total size of the population is small (e.g.\ the population is well below the carrying capacity or there is a high amount of resource relative to the number of bacteria). In such a case, both phenotypes can grow. Yet, we can still suppress the number of resistant bacteria and the average bacterial load over time. We explore the impact of various parameters and pulsed protocol durations on the average bacterial load over time. Our models also features stochasticity, which we develop in a stochastic kinetic framework \cite{wilkinson11}. We show that only when selection against the resistant type is high when the antibiotic is off can pulsed protocols effectively control the population.


\section{Methods}

Stochastic birth-death processes are widely used in biological modelling \cite{novozhilov06}, and, in particular, stochastic modelling of the Lotka-Volterra system \cite{huang15}. Our stochastic model features microscopic processes of birth, death, competition, and mutation, as detailed in Box \ref{stochasticLVprocesses}. These processes operate on two phenotypes $X$ and $Y$, which represent a wild-type strain, which is susceptible to the antibiotic, and a mutant strain, which is resistant, respectively.

Consider first the dynamics of the birth and death processes without the presence of antibiotics, i.e.\ in the antibiotic-free regime. Reaction set \ref{sp:birthdeath} represents these processes, where $b$ and $d$ are the birth and death rates, respectively, for the wild-type strain with $b>d>0$. We assume that the death rate for the resistant strain is also $d$. However, assuming a cost $c>0$ to the birth rate for resistance, the birth rate of the resistant strain is $b-c$ (costs applied to birth rather than death rates have also been applied similarly in ecological games \cite{hauert08}).

In the presence of the antibiotic, the above processes still occur, but with an additional set of reactions involving the antibiotic. Since we are considering a concentration-independent or time-dependent antibiotic, we will assume that the amount of antibiotic, $\bar{A}$, remains unchanged while we are in the antibiotic regime. At the maximum dose $\bar{A}=1$, normalized. The antibiotic is bactericidal, and kills both types of bacteria. 
Reaction set \ref{sp:antibiotic} represents death from the antibiotic with rates $\alpha$ and $\alpha'$ for the susceptible and resistant strains, respectively. Note that for $X$ to be susceptible and $Y$ to be resistant, we must have $\alpha>b-d>c+\alpha'$.

The species also die from competition as represented in Reaction set \ref{sp:comp}. Death occurs due to the bacteria competing for resources; the loser receives less and thus has some chance of dying. Let $\gamma$ be the rate at which two bacteria compete over a critical resource in a well-mixed population. We use the birth rate minus the death rate as a proxy for their competitiveness. We then assume that the rate of death of the focal type from competition is related to the ratio of their opponent's competitiveness to their own. i.e.\ since the wild-type is more competitive than the mutant, a mutant is more likely to die when competing for a resource with a wild-type bacterium. If the focal type and its competitor are the same type, then the ratio is equal to one. We define $\kappa$ as the rate of death of a mutant in competition with a wild-type. And, since competition may scale nonlinearly with respect to the ratio of competitiveness, we consider several values of $\kappa \geq 1$. We then assume that the rate of death of a wild-type in competition with a mutant is the inverse, i.e.\ $1/\kappa$. Frequently in Lotka-Volterra systems, intra-specific competition is assumed to be greater than inter-specific. As such, the rate of death would be greater for competition between bacteria of the same type than those that are different. However, we think it reasonable that between kin interactions can be greater than within kin interaction in our case, because the bacteria are well mixed.

The bacteria can turn into the other type via mutation at rate $\mu$, which occurs in both regimes (Reaction set \ref{sp:mutation}). However, under the stress of the antibiotic, the susceptible type $X$ will mutate to $Y$ at a higher rate $\mu'>\mu$.

\begin{mymathbox}[Stochastic Lotka-Volterra processes]{stochasticLVprocesses}
&\text{Birth/death: } & &X \xrightarrow{b} 2X & &Y \xrightarrow{b-c} 2Y & &X \xrightarrow{d} \emptyset & &Y \xrightarrow{d} \emptyset \label{sp:birthdeath} \\
&\text{Competition: } & &2X \xrightarrow{\gamma} X & &2Y \xrightarrow{\gamma} Y & &XY \xrightarrow{\gamma/\kappa} Y & &XY \xrightarrow{\gamma\kappa} X \label{sp:comp} \\
&\text{Death via antibiotic: } & &\bar{A}X \xrightarrow{\alpha} \emptyset & &\bar{A}Y \xrightarrow{\alpha'} \emptyset \label{sp:antibiotic} \\
&\text{Mutation: } & &X \xrightarrow{\mu(1-\bar{A})+\mu'\bar{A}} Y & &Y \xrightarrow{\mu} X
\label{sp:mutation}
\end{mymathbox}

The environmental switching is controlled by a choice of the on and off durations of the drug. We assume $100\%$ bioavailability of the drug at application, e.g.\ intravenous application. Thus, when the antibiotic is ``turned on,'' its effects are immediate. Further, when it is ``turned off,'' the dissipation of the antibiotic --- i.e.\ the rate at which it breaks into ineffective material, is metabolized, etc.\ --- is rapid, which is common for concentration-independent and time-dependent antibiotics. In the antibiotic regime, we apply the maximum effective dose. Thus, the set of pulsed protocols we consider are sequences of durations of the regimes, where $\bar{A} = 1$ when the antibiotic is on and $\bar{A} = 0$ when it is off.

\begin{table*}[!htpb]
\begin{center}
\begin{tabularx}{\textwidth}{ll}
\toprule
Parameter and default value & Definition \\
\midrule
$b = 0.35$ & Birth rate. \\
$c = 0.05$ & Cost for resistance. \\
$d = 0.1$ & Death rate. \\
$\alpha = 0.4$ & Death rate of the wild-type via antibiotic. \\
$\alpha' = 0.1\alpha = 0.04$ & Death rate of the mutant via antibiotic. \\
$\gamma = 10^{-5}$ & Competition rate. \\
$\kappa>1$ & Death rate of mutants from competition with the wild-type. \\
$\mu = 10^{-5}$ & Mutation rate. \\
$\mu' = 10\mu = 10^{-4}$ & Stress induced mutation rate of susceptible to resistant bacteria. \\
\bottomrule
\end{tabularx}
\end{center}
\vspace{-2mm}
\caption{Summary definitions of parameters and variables.} \label{param}
\end{table*}

We conducted numerical simulations of the model to test the effects of various protocols. We average realizations for each parameter combination. We use the Gillespie algorithm \cite{gillespie76} from Julia's DifferentialEquations and Catalyst packages \cite{rackauckas17}. Table \ref{param} lists the default parameter values used with rates per $15$ minutes. We vary these values to explore nearby parameter space. 
We assume that a new generation occurs after $1/(b-d) = 1$ hour. We estimate that the relative fitness of the resistant strain in the antibiotic-free environment is $(b-c-d)/(b-d) \approx 0.8 \implies c=0.05$, which is within experimentally evaluated values \cite{melnyk15}. \textit{E.\ Coli} has an average rate of mutation per genome per generation on the order of $10^{-5}$ \cite{perfeito07}. Further, under stress from the antibiotic, the mutation rate can be even larger, up to ten times the non-stressed rate \cite{kuban04}. Thus, we consider $\mu'=10\mu$. We assume that resistant bacteria die from the antibiotic at $1/ \nth{10}$ the rate susceptibles do \cite{coates18}, i.e.\ $\alpha'=0.1\alpha$. The initial condition is a population of $100\%$ susceptible bacteria, $X_0 = 10^5$. We explore a variety of competition parameters $\kappa$. We consider fixed on/off durations, where we repeat switching until the population is extinct or $t=14$ days.

\section{Results}

We first consider the mean field (i.e. non-stochastic) behaviour of the system with average values $\m{X}$ and $\m{Y}$ (for details of the derivation see Appendix \ref{app:derivation}.)
\begin{align}
\dot{\m{X}} &= (b - d - \mu(1-\bar{A}) - (\alpha+\mu')\bar{A})\m{X} + \mu\m{Y} - \gamma\m{X}^2 - \frac{\gamma}{\kappa}\m{X}\m{Y}, \\
\dot{\m{Y}} &= (b - c - d - \mu - \alpha'\bar{A})\m{Y} + (\mu(1-\bar{A}) + \mu'\bar{A})\m{X} - \gamma\kappa\m{X}\m{Y} - \gamma\m{Y}^2.
\end{align}
\noindent In either environment, both types will coexist at equilibrium due to mutations, though the less adapted type will remain at low frequency. When the population size is sufficiently low (i.e.\ below the zero isoclines), competition will also be low, which will allow both types to increase in abundance. However, the higher the competition term $\kappa$, the smaller this region is. A large $\kappa$ will cause the mutant strain to be suppressed in the antibiotic-free environment, and in the stochastic scenario, this effect increases the chance that the mutant strain be eliminated. In the remainder of the results, we detail the effects of switching the drug on and off, competition, and stochasticity in the stochastic scenario.

\begin{figure}[!ht]
\centering
\begin{subfigure}[A]{\textwidth}
    \caption{}\label{fig:tsOn_fail}
    \includegraphics[width=\textwidth]{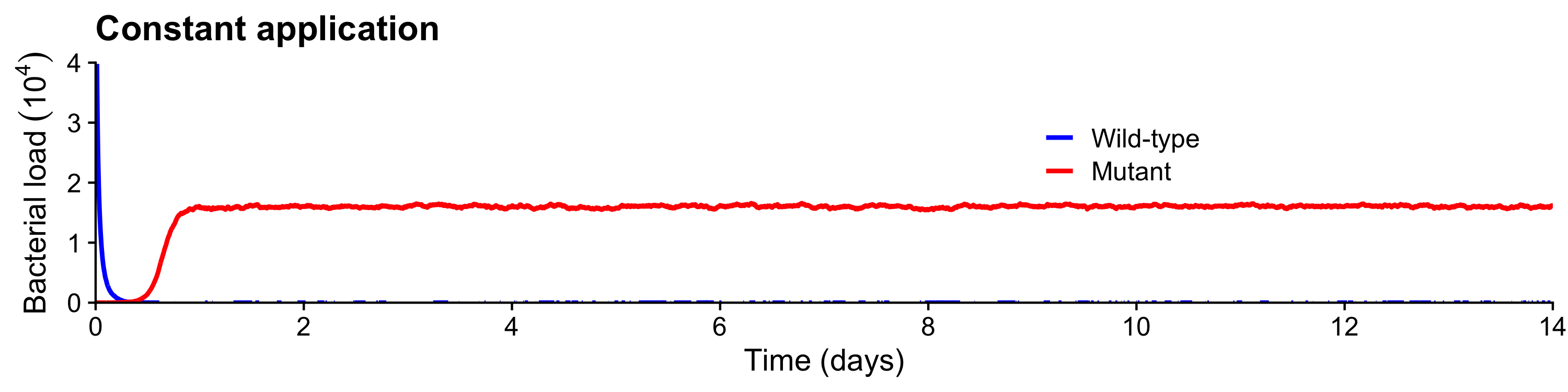}
\end{subfigure} \\
\begin{subfigure}[A]{\textwidth}
    \caption{}\label{fig:tsPulse_succ}
    \includegraphics[width=\textwidth]{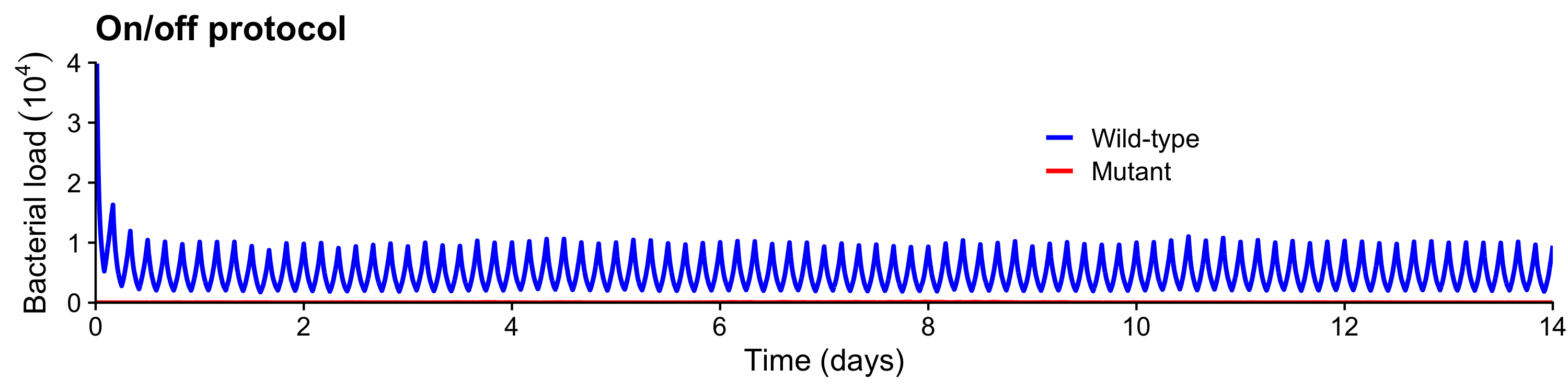}
\end{subfigure} \\
\begin{subfigure}[A]{\textwidth}
    \caption{}\label{fig:tsPulse_fail_1}
    \includegraphics[width=\textwidth]{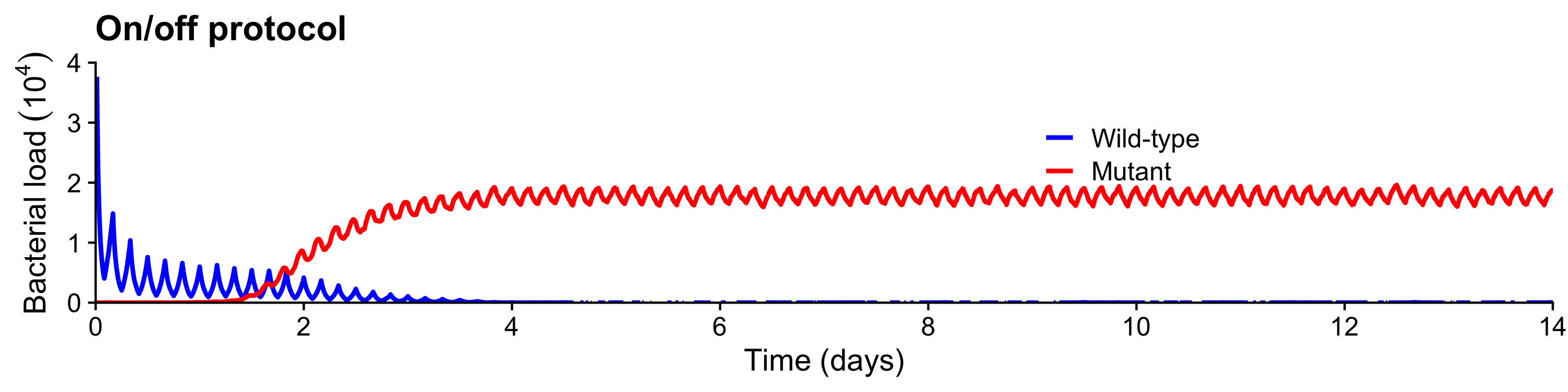}
\end{subfigure} \\
\begin{subfigure}[A]{\textwidth}
    \caption{}\label{fig:tsPulse_fail_2}
    \includegraphics[width=\textwidth]{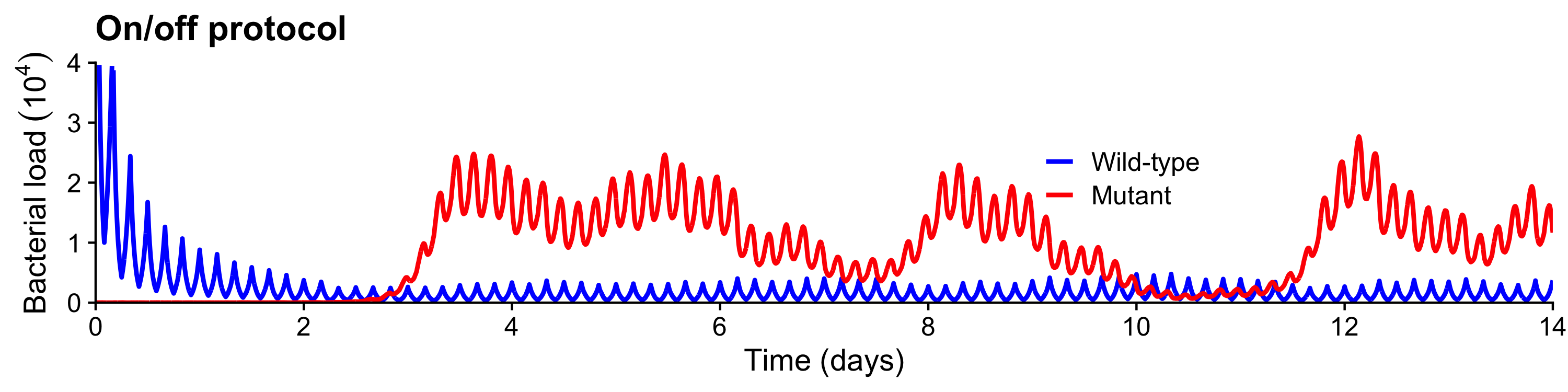}
\end{subfigure}
\caption{Representative times series for resistance emerging under constant application, panel (\text{a}), and results for a pulsed protocol (on and off for $2$hrs each), panels (\text{b})-(\text{d}). Blue are wild strain and red resistant. Pulsed protocols can suppress the bacterial load (\textbf{b} with $\alpha=0.4$ and $\kappa=4$). However, if we increase the antibiotic kill rate to $\alpha=0.45$, the pulsed protocol fails (\textbf{c}). We may also observe oscillations (\textbf{d} with $\alpha=0.5$, $\gamma=0.5 \times 10^{-5}$, $\kappa=20$, and $\mu=10^{-4}$). $X_0=10^5$ wild-type strain bacteria, and the remaining parameters, if not mentioned here, are from Table \ref{param}. Note that we only depict results $\leq 10^4$, since the population is quickly contained to this region.}
\label{fig:LVts}
\end{figure}

Figure \ref{fig:LVts} depicts a representative time series for a switching protocol vs.\ a constant application of the antibiotic. With a sufficient competitive disadvantage for resistance, i.e.\ high $\kappa$, we can effectively suppress the average bacterial load over time and resistant bacteria relative to a constant application as depicted in Figures \ref{fig:tsPulse_succ} (pulsed protocol) and \ref{fig:tsOn_fail} (constant application). However, due to stochastic effects, resistance can briefly rise as seen in Figure \ref{fig:tsPulse_fail_2} before it is brought under control again. This phenomenon emerges from the interactions between the two types at different time scales. Oscillations between the types emerge at a time scale longer than the period of the protocol. When the mutants are at their peak, the wild-type strain is not, and vice verse. However, peaks of the wild-type strain are relatively low as can be seen in panel d. Between day $10$ until sometime during day $12$, the mutants are suppressed and the wild-type are at a relative high. This high, however, is not much higher than when they are suppressed. As such, the total bacterial load is relatively low. Regardless of these waves of resistant bacteria, the average bacterial load over time can still be less than when the mutant becomes established under constant application, as we shall show.

Though Figure \ref{fig:tsOn_fail} depicts a specific instance where the mutant becomes established under the constant application protocol, resistance can be prevented by rapid elimination of the population. Since the mutations and fluctuations in abundances are stochastic, it is possible that we are fortunate and constant application drives the population extinct before resistance emerges. Thus, to better understand the effectiveness of therapies we must evaluate the statistics of the bacterial load as a function of system parameters. We will show that constant application of an antibiotic tends to lead to either extreme: elimination of the entire population, or the establishment of a majority resistant population.

Averaged over $50$ realizations, we calculate average bacterial load over time for pulsed protocols with various on and off durations and compare these to the bacterial load for constant application of the antibiotics. We plot these results in heat maps, where the colour indicates the long-term bacterial load relative to the outcome from constant antibiotic application. Red indicates that the therapy is on average worse than constant application, yellow is on average equal, and blue indicates that it is on average better.

We observe that pulsed protocols along a diagonal do best. One reason for this is that the switching times explored here are much less than the time to reach carrying capacity in either regime. For example, even a day-period protocol will not reach carrying capacity (the expected time to reach carrying capacity is between one and two days). In such a case, the population can swing from predominantly one type to the other (see Appendix \ref{app:furthersims} for an example time series). However, this behaviour can still be beneficial, since each application of the antibiotic is another chance of eliminating the population, since switching environments drives the dominant type down potentially to extinction before the other type can become established. In addition to plotting heat maps, we plot the average bacterial load over time for constant and pulsed ($2$hrs on and off each) for various values of each parameter averaged over $50$ realizations.

\begin{figure}[!htb]
    \centering
    \begin{subfigure}[]{0.3\textwidth}
        \caption{}
        \includegraphics[width=\textwidth]{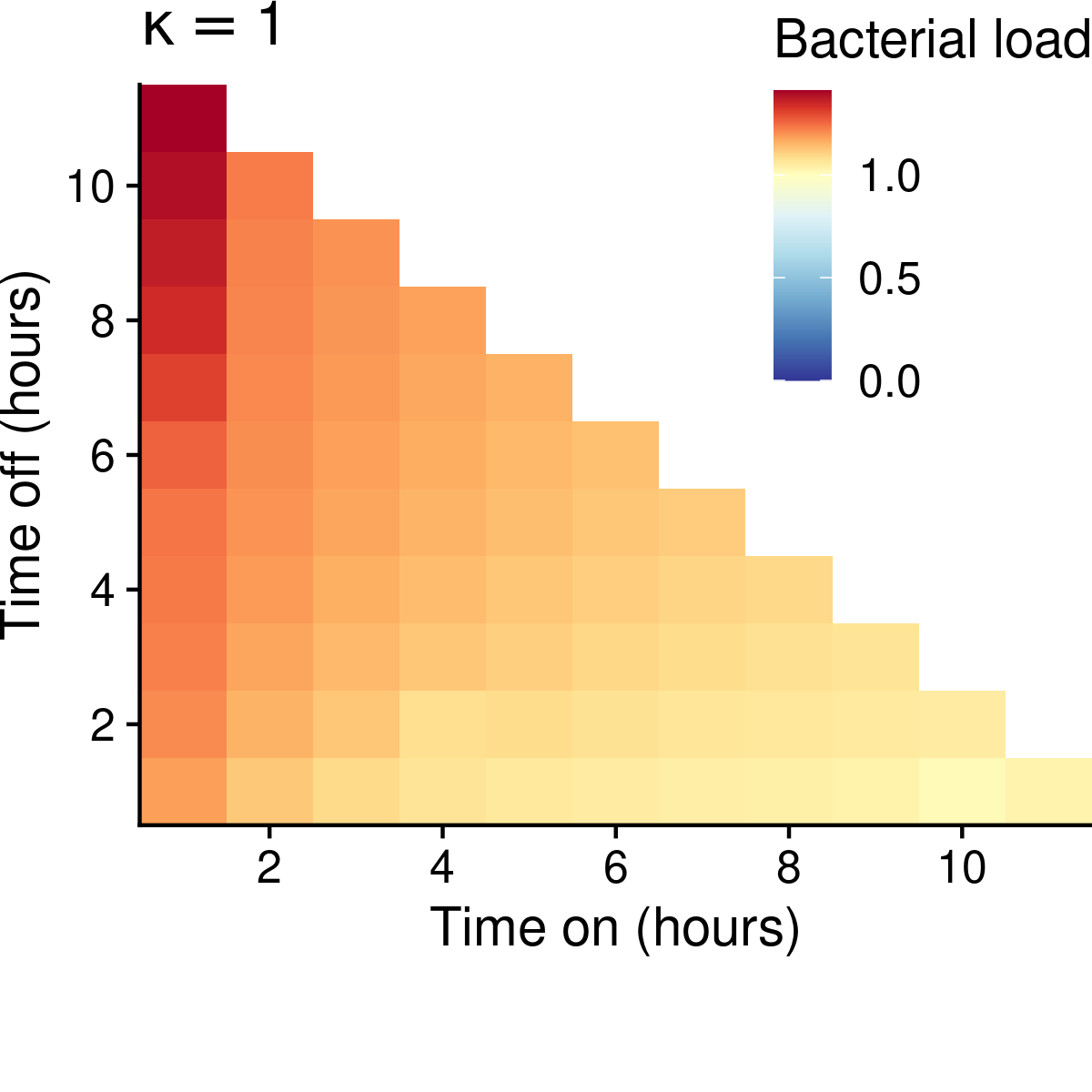}
    \end{subfigure}
    \begin{subfigure}[]{0.3\textwidth}
        \caption{}
        \includegraphics[width=\textwidth]{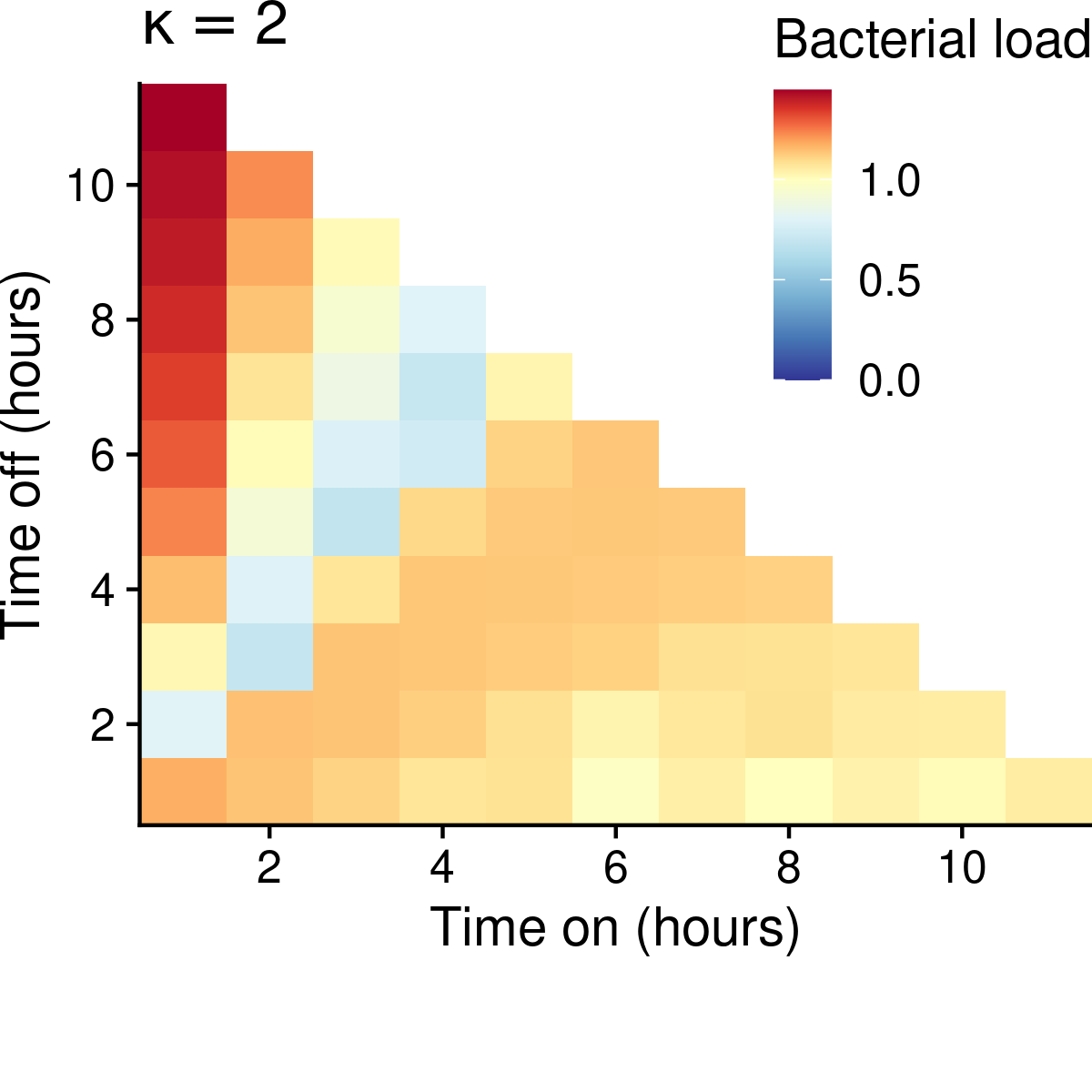}
    \end{subfigure}
    \begin{subfigure}[]{0.3\textwidth}
        \caption{}
        \includegraphics[width=\textwidth]{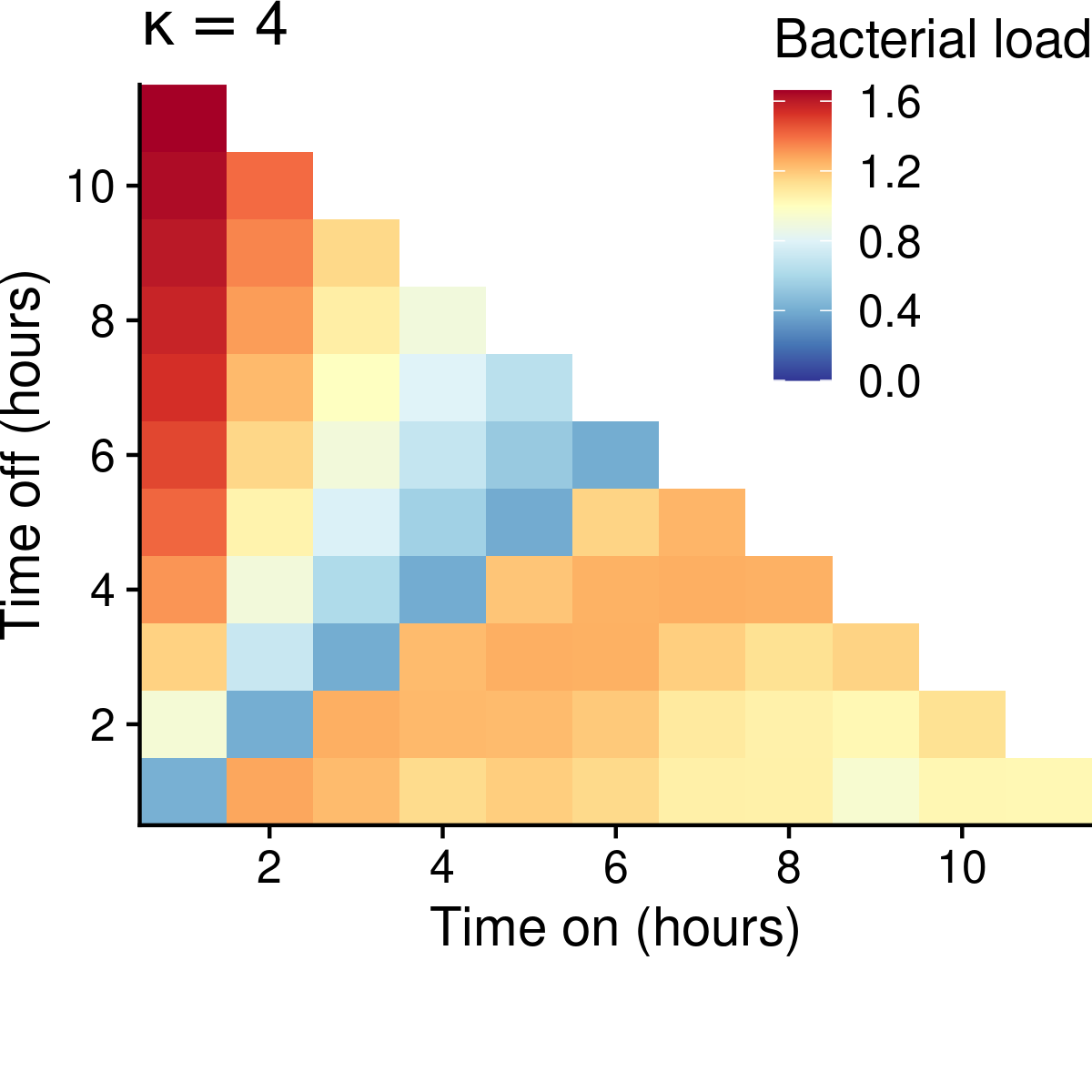}
    \end{subfigure} \\
    \begin{subfigure}[]{0.3\textwidth}
        \caption{}
        \includegraphics[width=\textwidth]{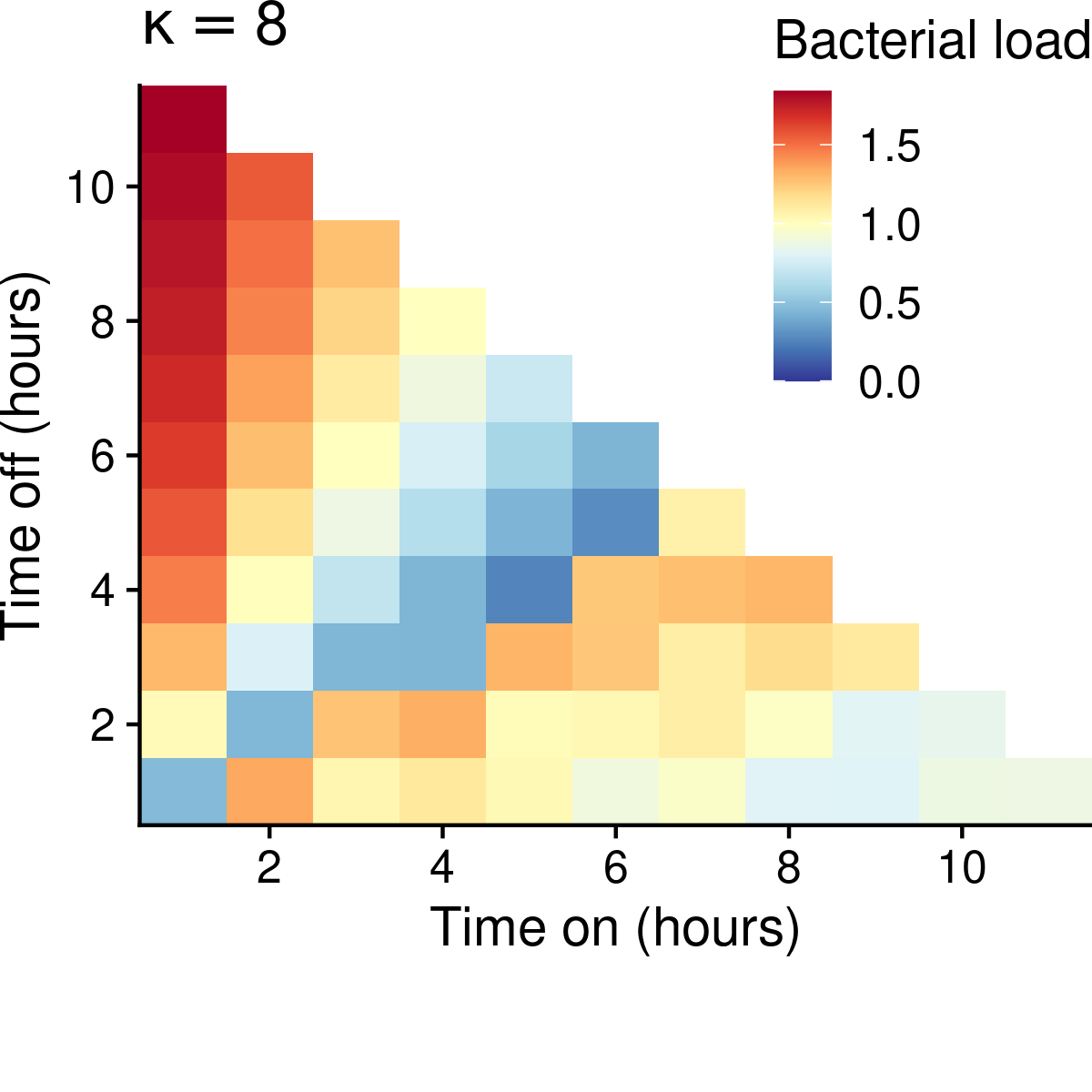}
    \end{subfigure}
    \begin{subfigure}[]{0.3\textwidth}
        \caption{}
        \includegraphics[width=\textwidth]{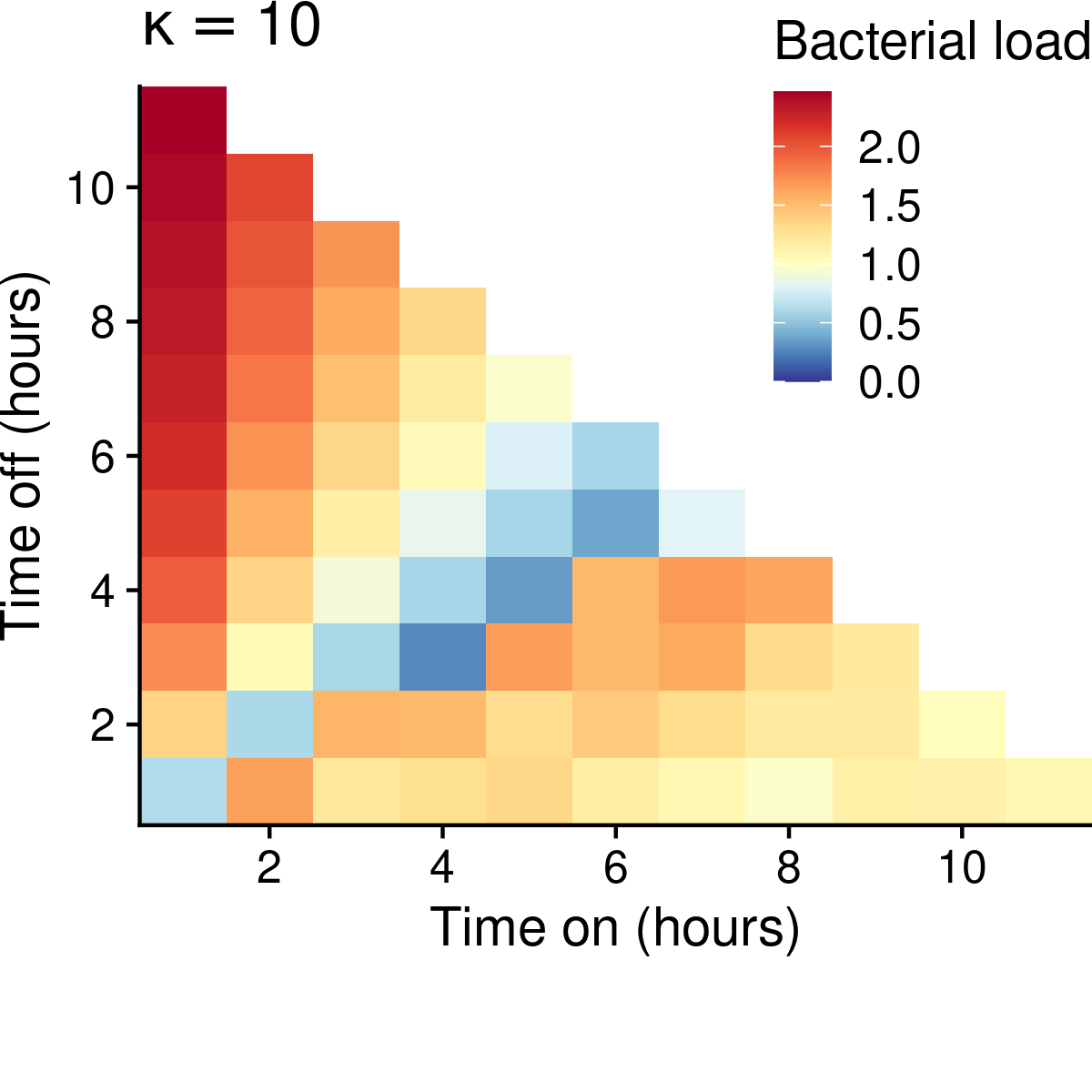}
    \end{subfigure}
    \begin{subfigure}[]{0.3\textwidth}
        \caption{}
        \includegraphics[width=\textwidth]{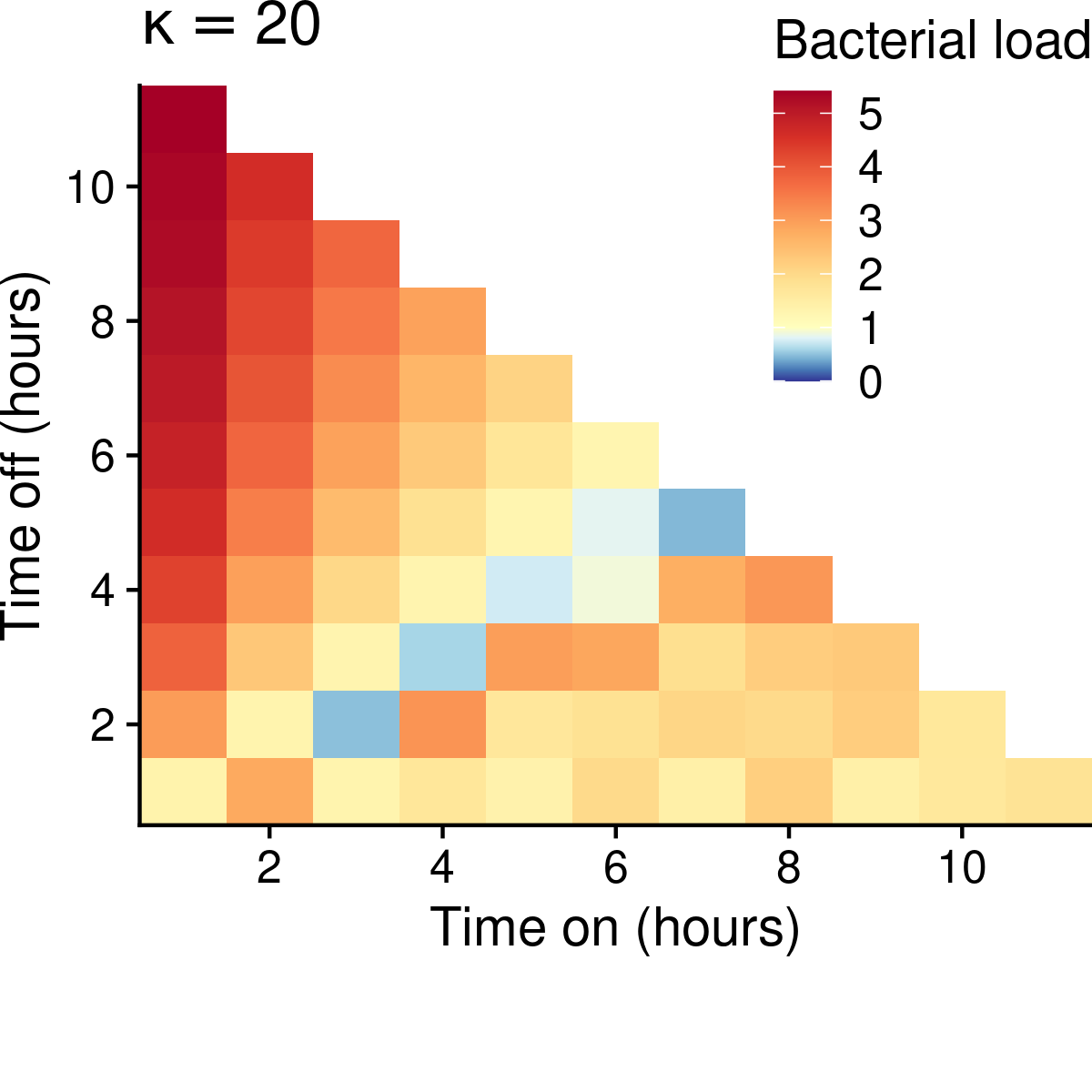}
    \end{subfigure} \\
    \begin{subfigure}[]{0.9\textwidth}
        \caption{}
        \includegraphics[width=\textwidth]{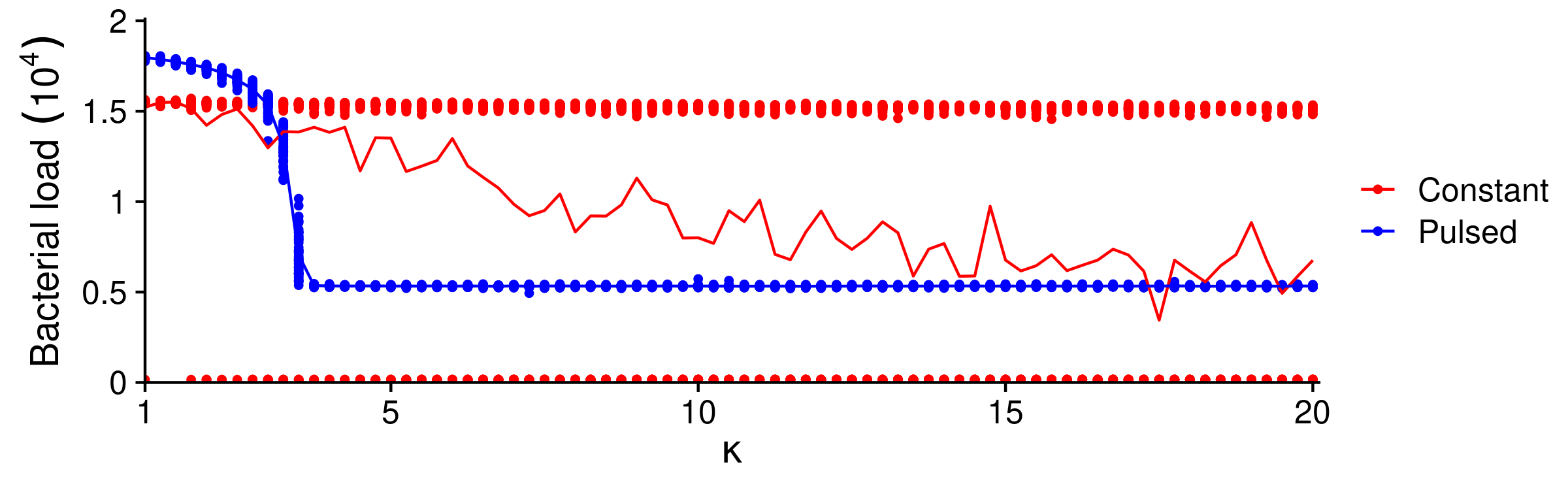} \label{fig:kappa_g}
    \end{subfigure}
    \caption{Heatmaps of the average bacterial load over time from pulsed protocols relative to that of constant application of the antibiotic for $\kappa= 1, 2, 4, 8, 10, 20$. Protocols that matched the average outcome of the constant application therapy are coloured in yellow. Those protocols that did worse are in red, and those that did better are in blue. Panel \textbf{g} depicts the average bacterial load over time for constant and pulsed ($2$hrs on and off each) for various $\kappa$. The points are the results for individual realizations and the curves their average.}
    \label{fig:kappa}
\end{figure}

Figure \ref{fig:kappa} shows that the higher the competition, the lower the diagonal (i.e.\ the best results come from protocols where the duration on is greater than the duration off). The increased competition suppresses the emergence of resistance even in the antibiotic environment, and thus the duration of application can be longer. However, we note that the effectiveness of the optimal pulsed protocols drastically falls if the duration on relative to off is too high. We do not see this sensitivity when reducing the time on. We also observe an intermediate level of competition is best for pulsed protocols relative to constant application. We can see this effect in Figure \ref{fig:kappa_g}. Increasing $\kappa$ decreases the average of the bacterial load of the individual realizations for the constant application as we would expect. Since, high competition between the types will suppress the emergence of resistant mutants (which is true in both environments). However, increasing competition has an initially steeper effect upon pulsed protocols before it levels off. A sufficient amount of competition is required for pulsed protocols to work. As $\kappa$ is increased, the difference between the outcomes of the two approaches decreases.

\begin{figure}[!htb]
    \centering
    \begin{subfigure}[]{0.3\textwidth}
        \caption{}
        \includegraphics[width=\textwidth]{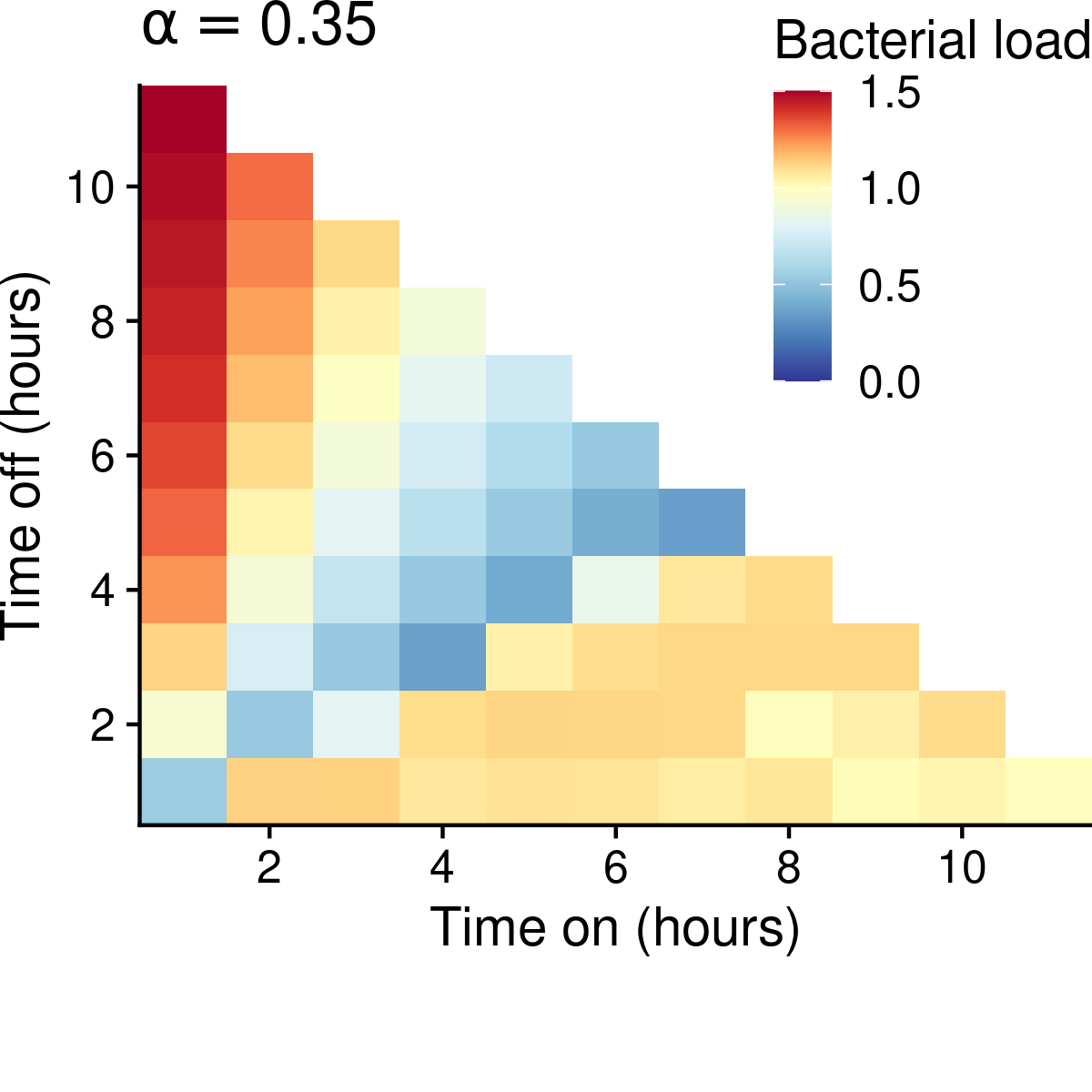}
    \end{subfigure}
    \begin{subfigure}[]{0.3\textwidth}
        \caption{}
        \includegraphics[width=\textwidth]{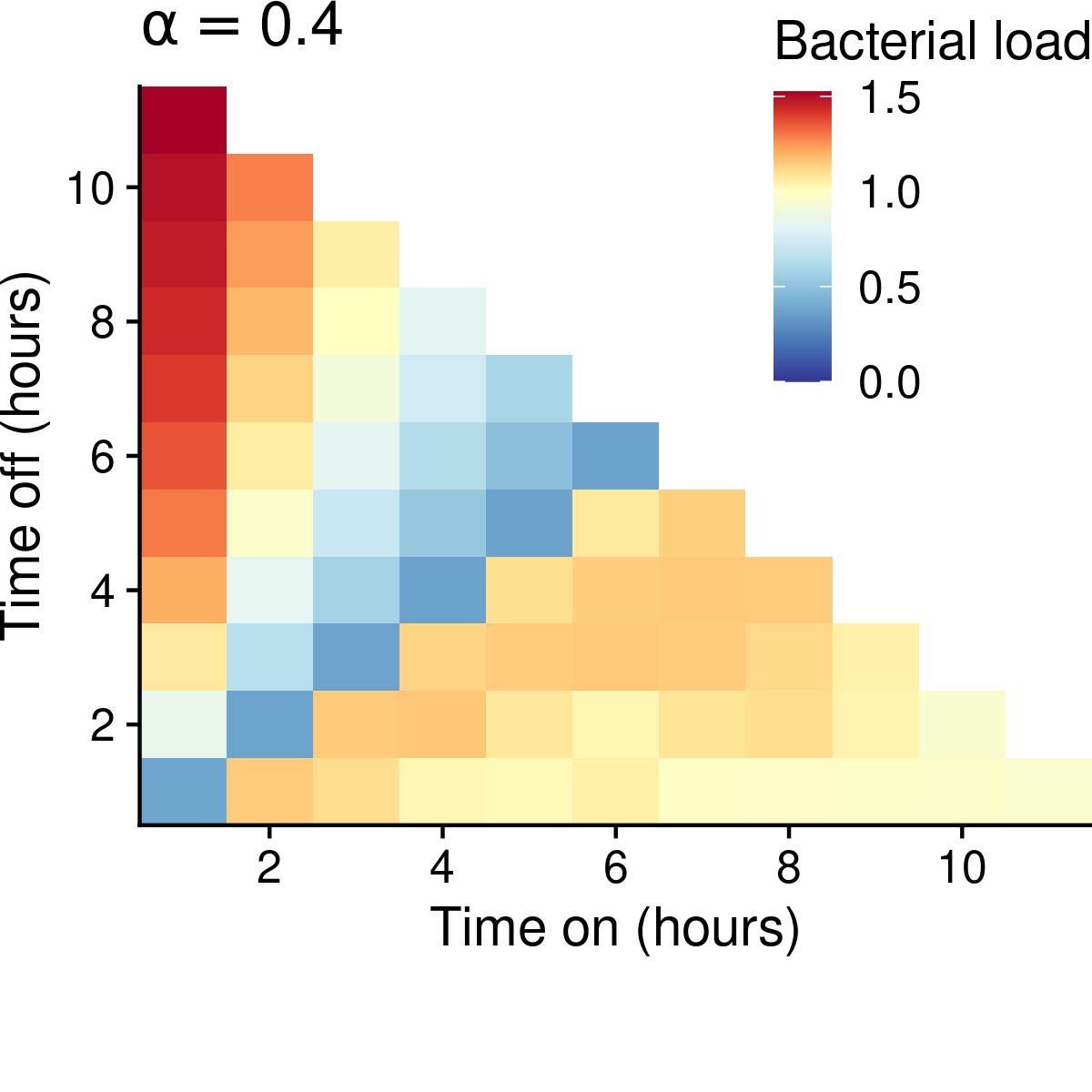}
    \end{subfigure}
    \begin{subfigure}[]{0.3\textwidth}
        \caption{}
        \includegraphics[width=\textwidth]{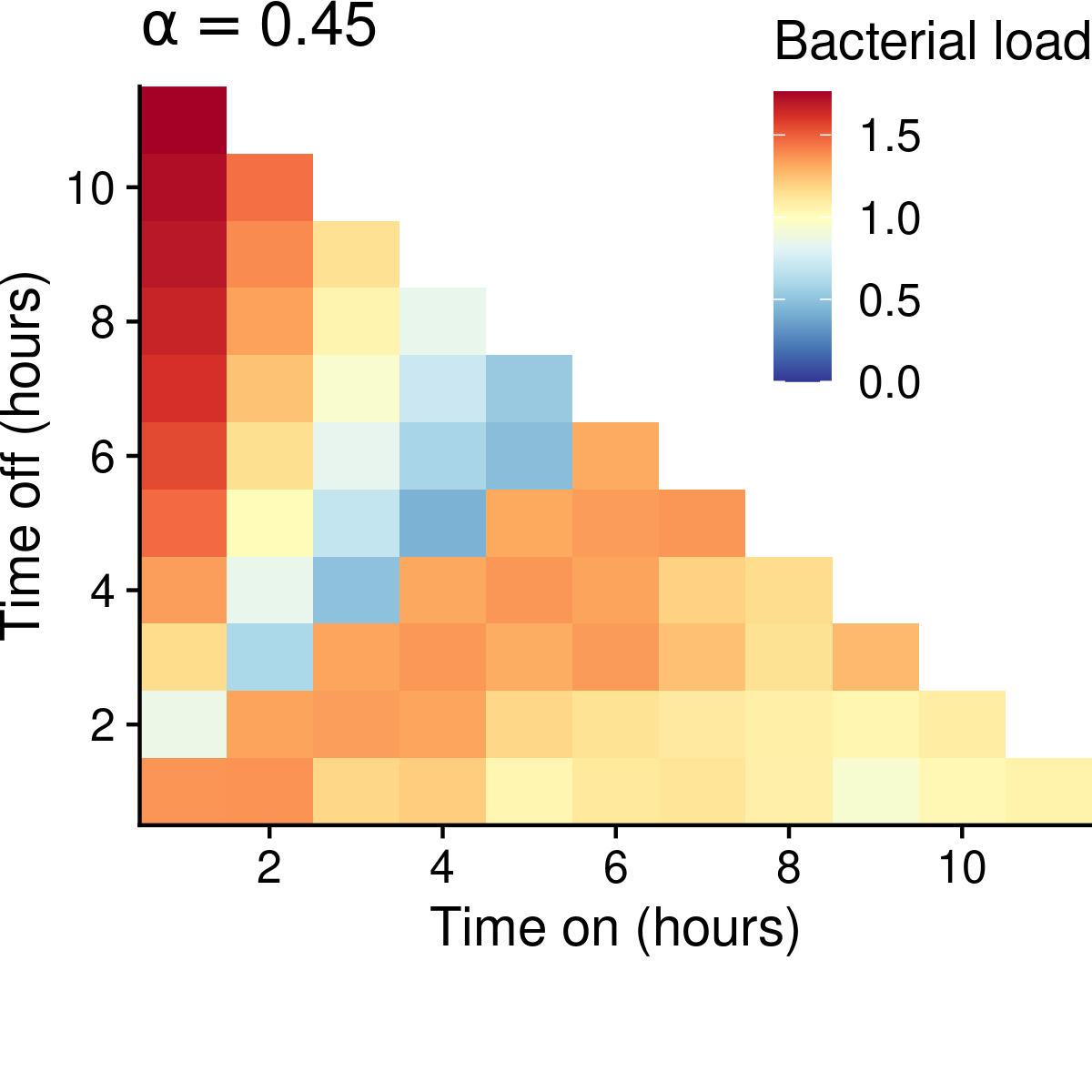}
    \end{subfigure} \\
        \begin{subfigure}[]{0.3\textwidth}
        \caption{}
        \includegraphics[width=\textwidth]{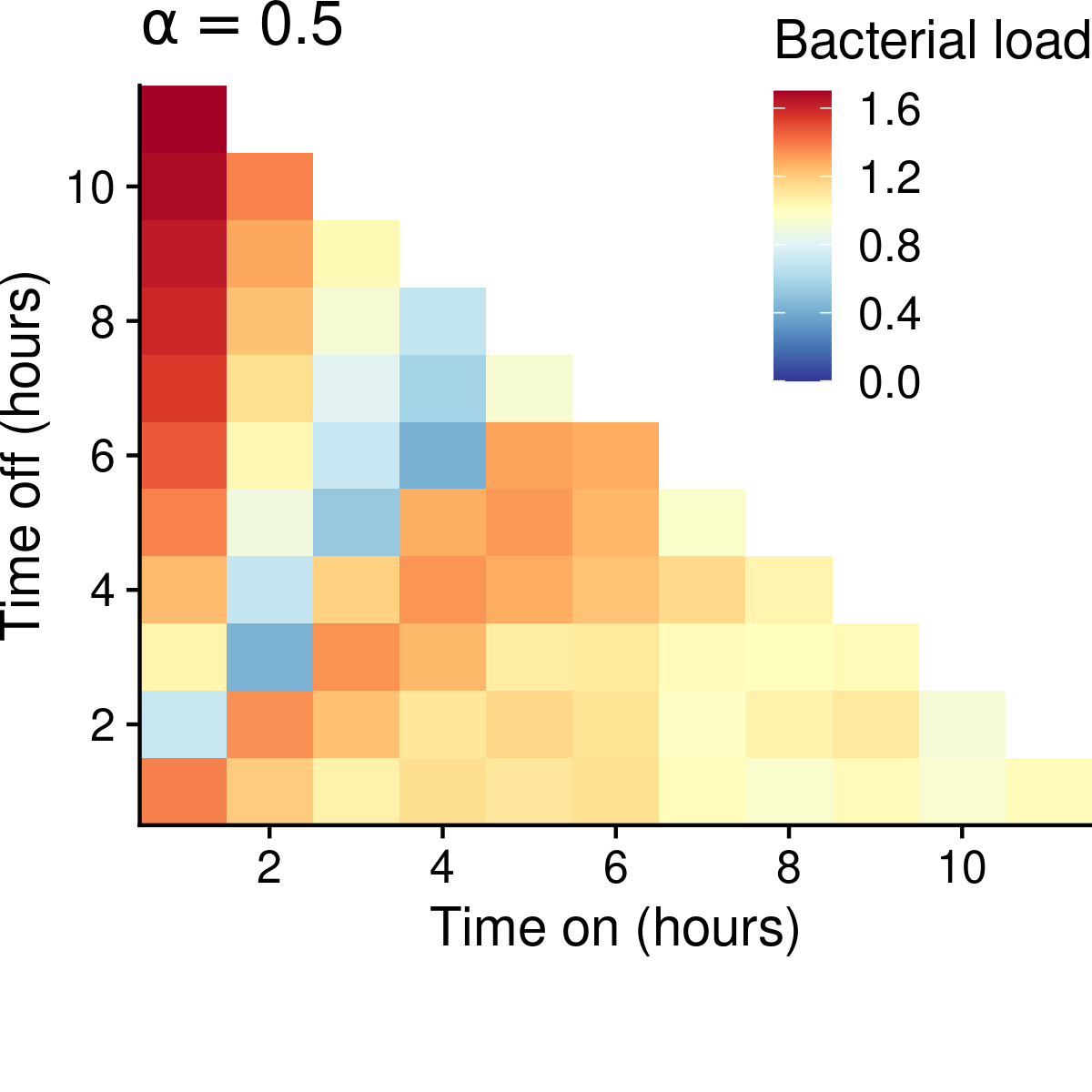}
    \end{subfigure}
    \begin{subfigure}[]{0.3\textwidth}
        \caption{}
        \includegraphics[width=\textwidth]{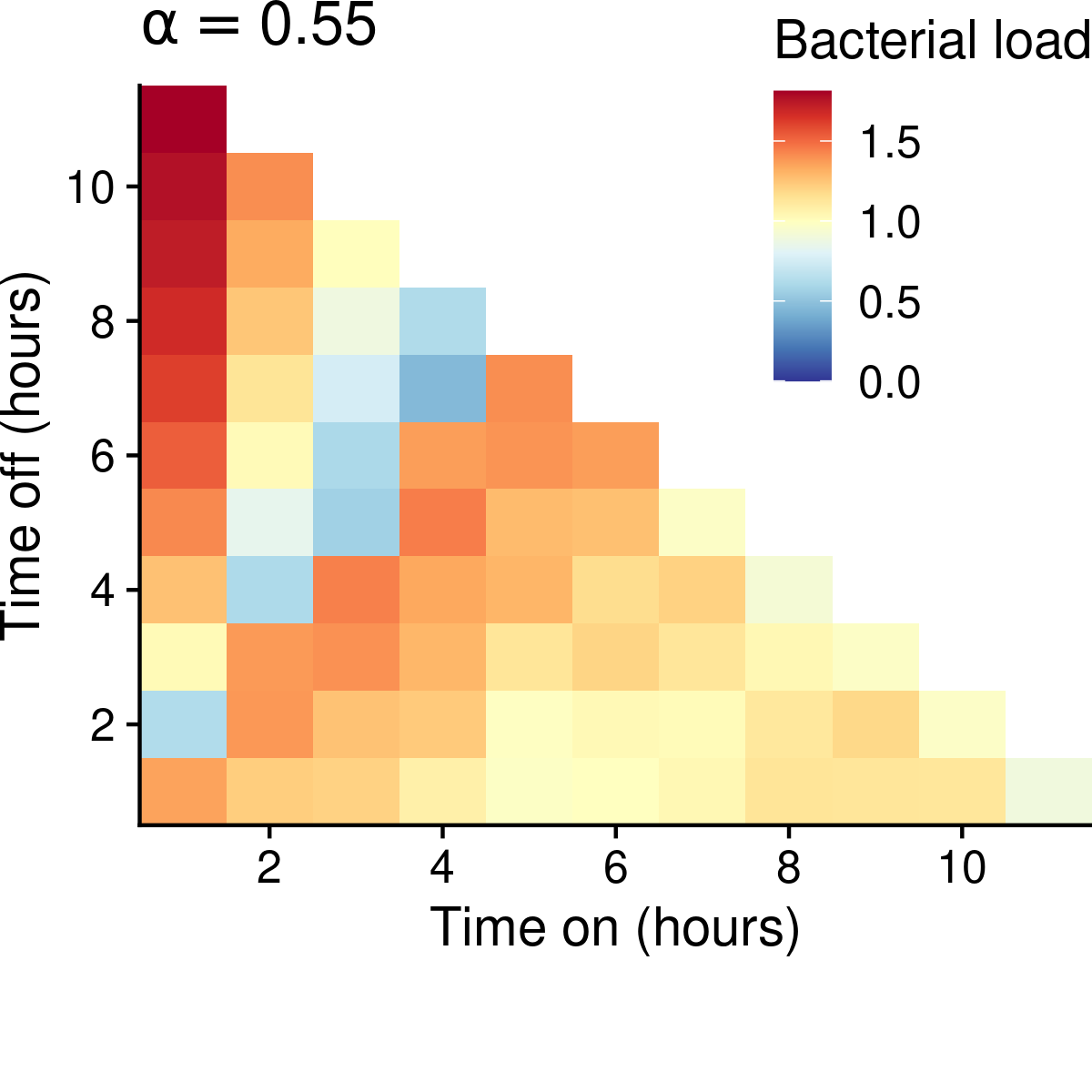}
    \end{subfigure}
    \begin{subfigure}[]{0.3\textwidth}
        \caption{}
        \includegraphics[width=\textwidth]{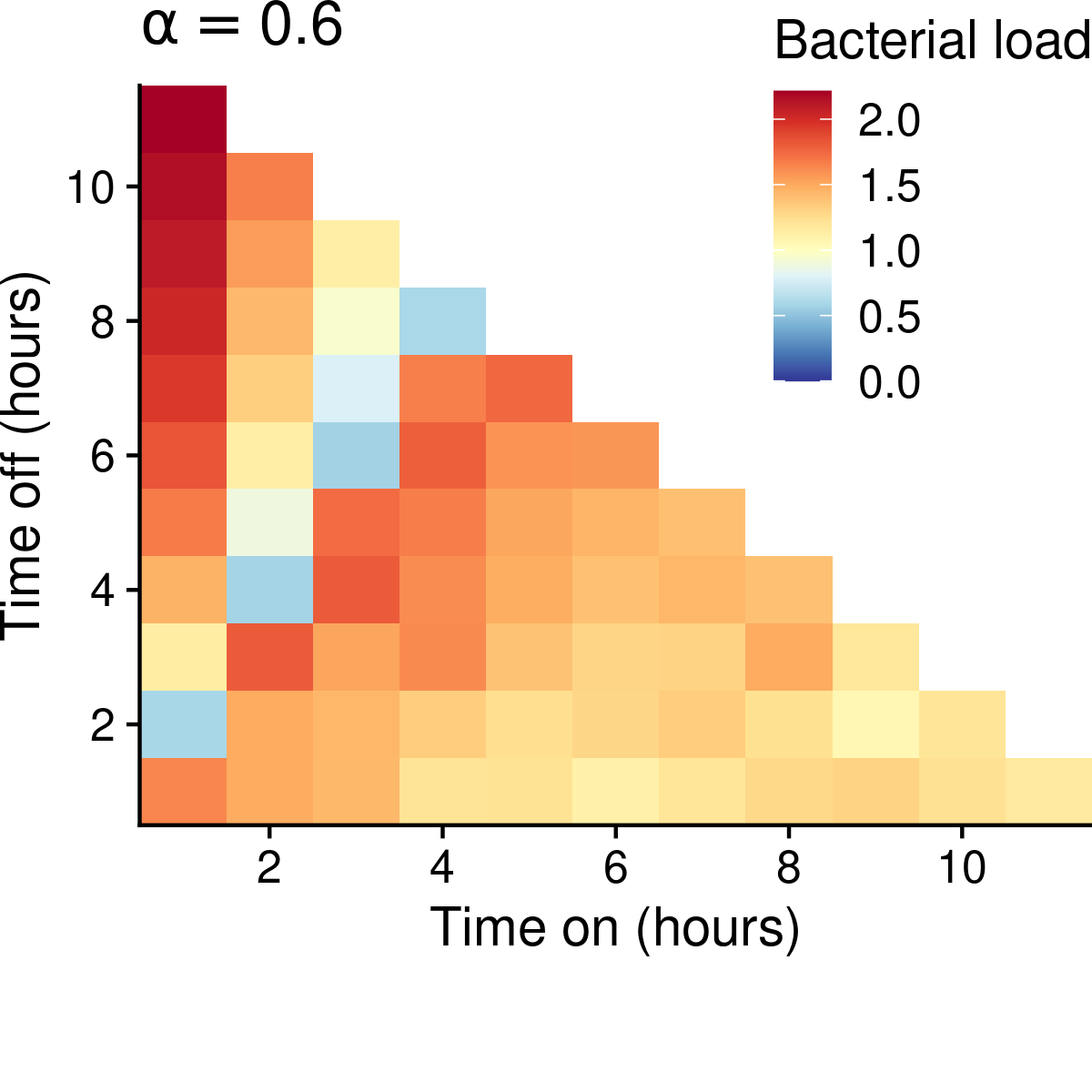}
    \end{subfigure} \\
    \begin{subfigure}[]{0.9\textwidth}
        \caption{}
        \includegraphics[width=\textwidth]{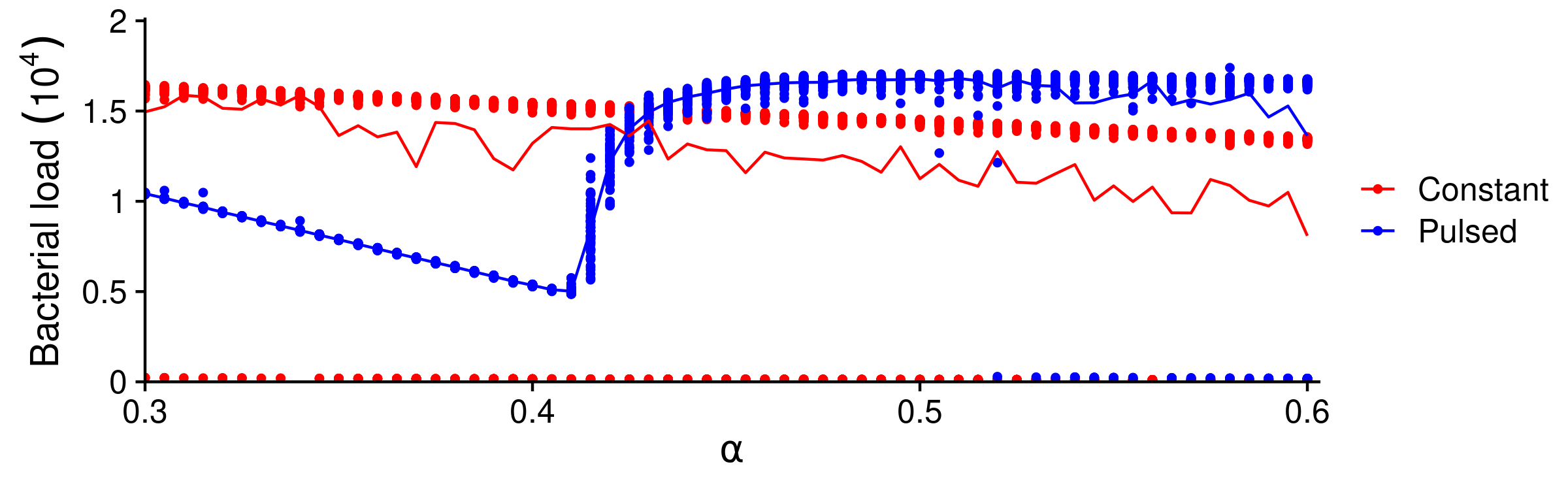} \label{fig:alpha_g}
    \end{subfigure}
    \caption{Heatmaps of the average bacterial load over time from pulsed protocols relative to that of constant application of the antibiotic for $\alpha = 0.35,0.4,0.45,0.5,0.55,0.6$ and $\kappa=4$ (panels \textbf{a}-\textbf{f}). Protocols that matched the average outcome of the constant application therapy are coloured in yellow. Those protocols that did worse are in red, and those that did better are in blue. Panel \textbf{g} depicts the average bacterial load over time for constant and pulsed ($2$hrs on and off each) for various $\alpha$. The points are the results for individual realizations and the curves their average.}
    \label{fig:alpha}
\end{figure}

Another reason for angle of the optimal diagonal of successful protocols in Figure \ref{fig:kappa}, is due to the relationships between the mean growth rates and the antibiotic kill rates. Figure \ref{fig:alpha} depicts the results for various values of $\alpha$. The higher the antibiotic kill rate, the shorter the duration on for the most successful protocols. Like in the case of $\kappa$, $\alpha$ impacts the effectiveness of pulsed protocols nonlinearly. Figure \ref{fig:alpha_g} depicts the mean results for various $\alpha$. The higher the $\alpha$, the better constant application does. However, this is not true for pulsed protocols. An intermediate value is best. This result is due to the impact of $\alpha$ on competitiveness. If $\alpha$ is too high, then the wild-type is suppressed too much, and thus cannot be used to suppress the mutant strain through competition. Figure \ref{fig:tsPulse_fail_1} depicts a time series of the case where $\alpha$ is too high, resulting in failure of the pulsed protocol. Increasing competition $\kappa$, however, can mitigate this effect, shifting the minimum to the right (see Appendix \ref{app:furthersims} for an example).

\begin{figure}[!htb]
    \centering
    \begin{subfigure}[]{0.3\textwidth}
        \caption{}
        \includegraphics[width=\textwidth]{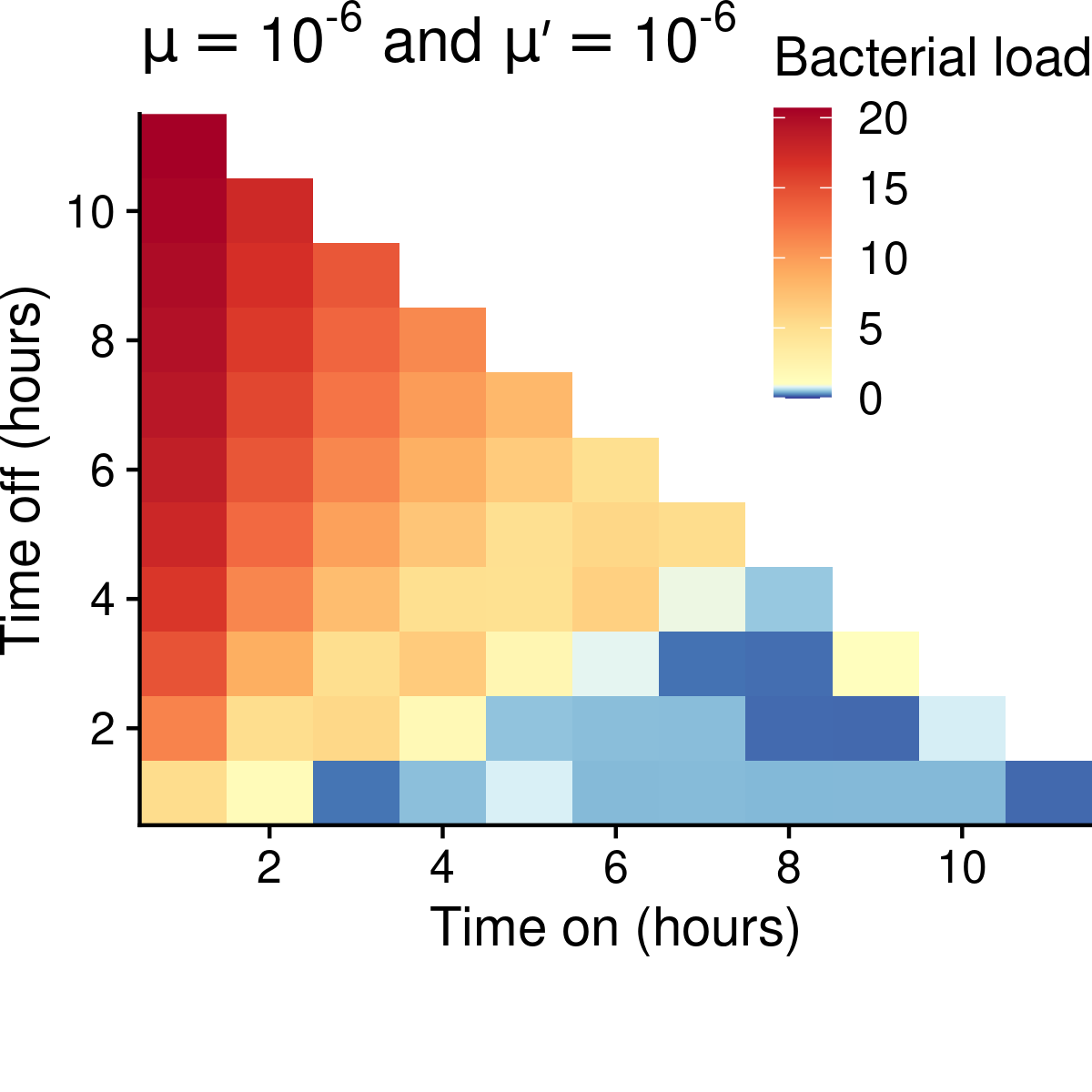}
    \end{subfigure}
    \begin{subfigure}[]{0.3\textwidth}
        \caption{}
        \includegraphics[width=\textwidth]{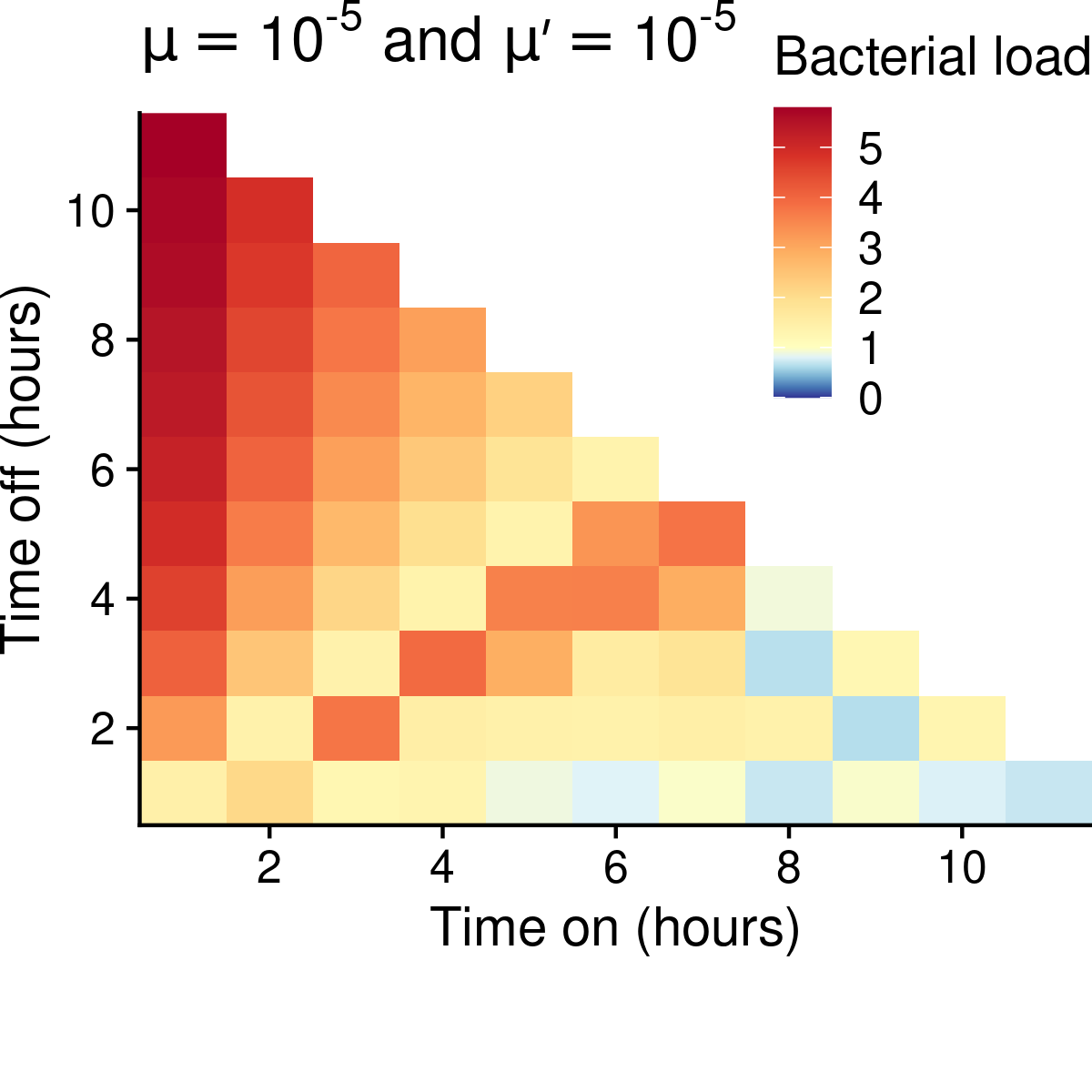}
    \end{subfigure}
    \begin{subfigure}[]{0.3\textwidth}
        \caption{}
        \includegraphics[width=\textwidth]{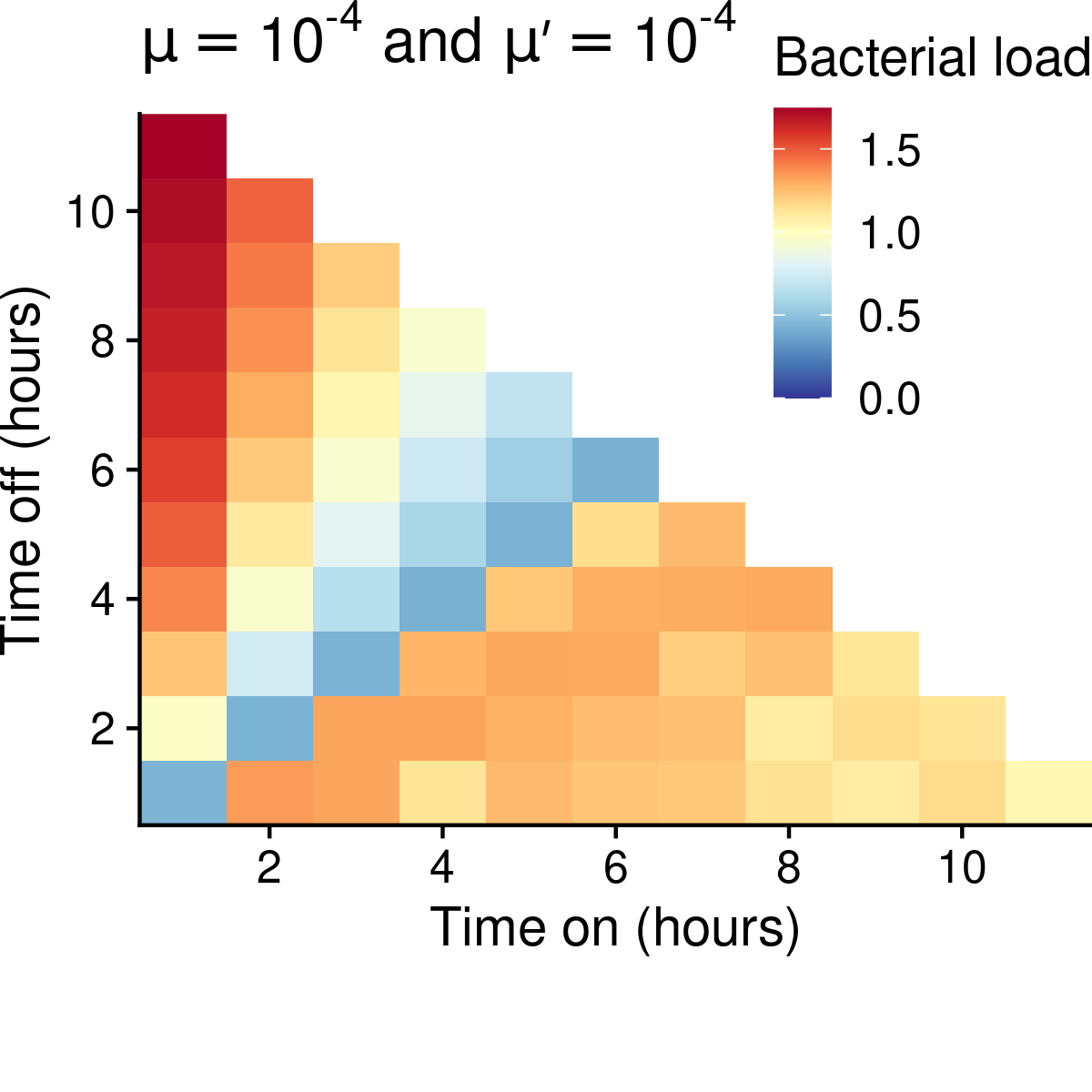}
    \end{subfigure} \\
    \begin{subfigure}[]{0.3\textwidth}
        \caption{}
        \includegraphics[width=\textwidth]{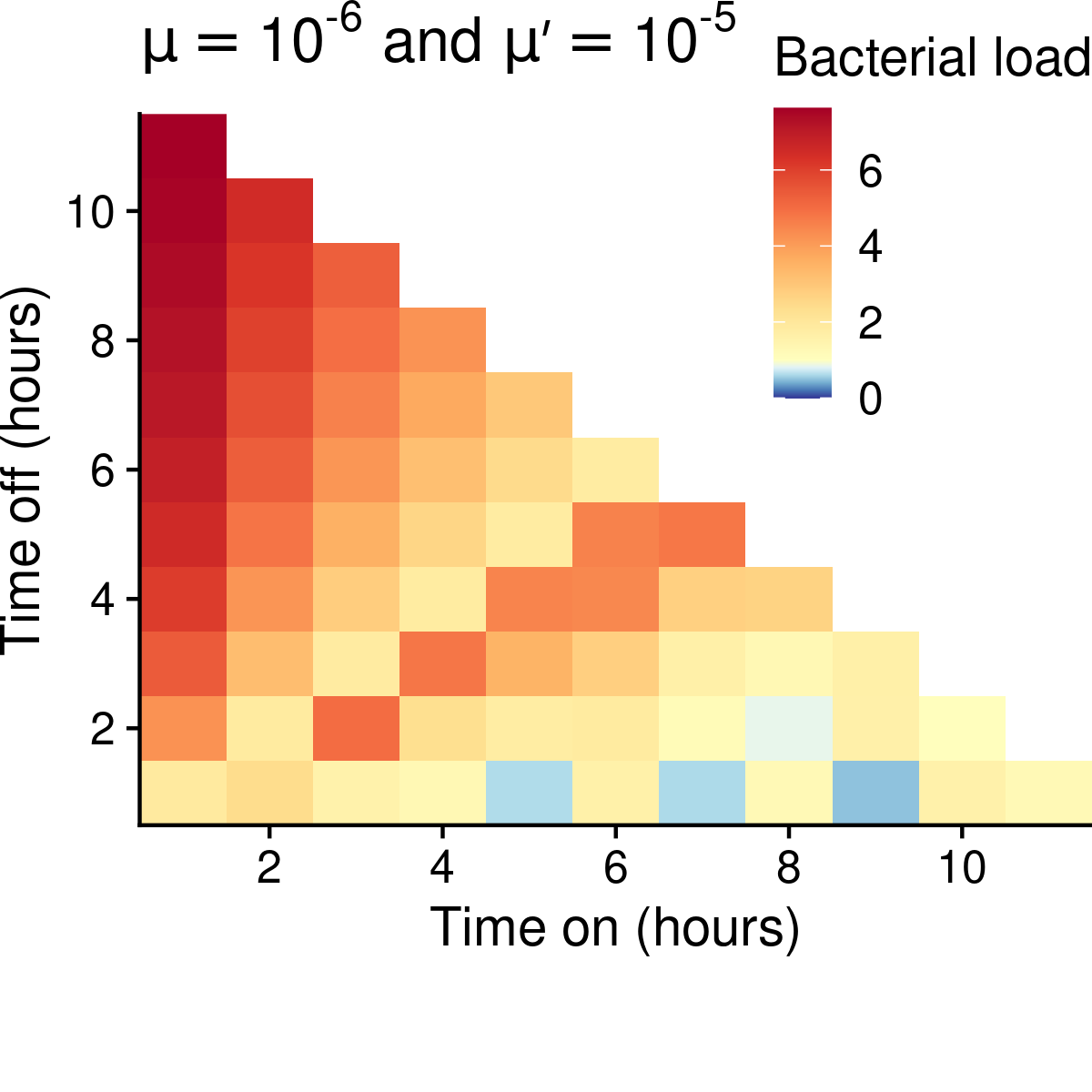}
    \end{subfigure}
    \begin{subfigure}[]{0.3\textwidth}
        \caption{}
        \includegraphics[width=\textwidth]{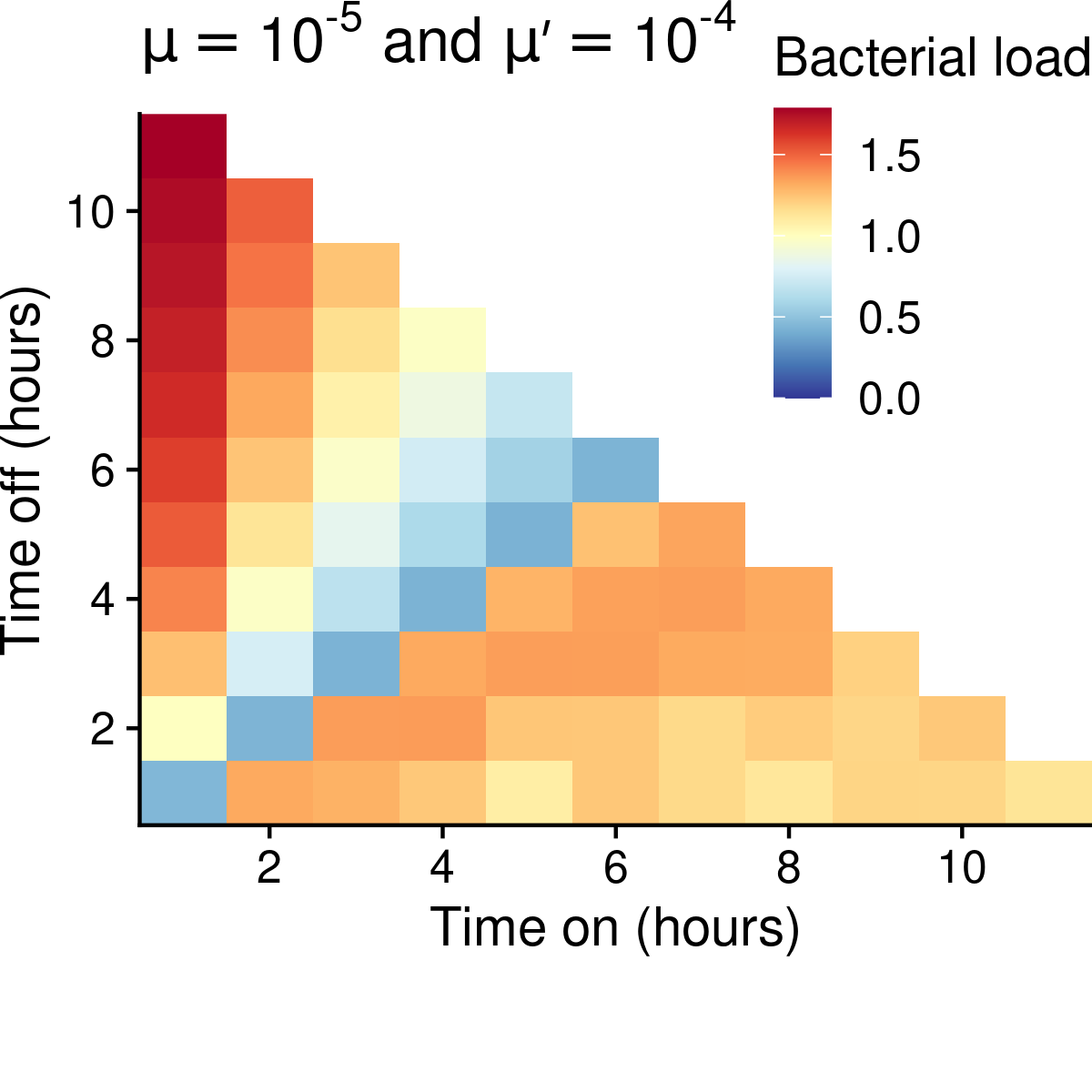}
    \end{subfigure}
    \begin{subfigure}[]{0.3\textwidth}
        \caption{}
        \includegraphics[width=\textwidth]{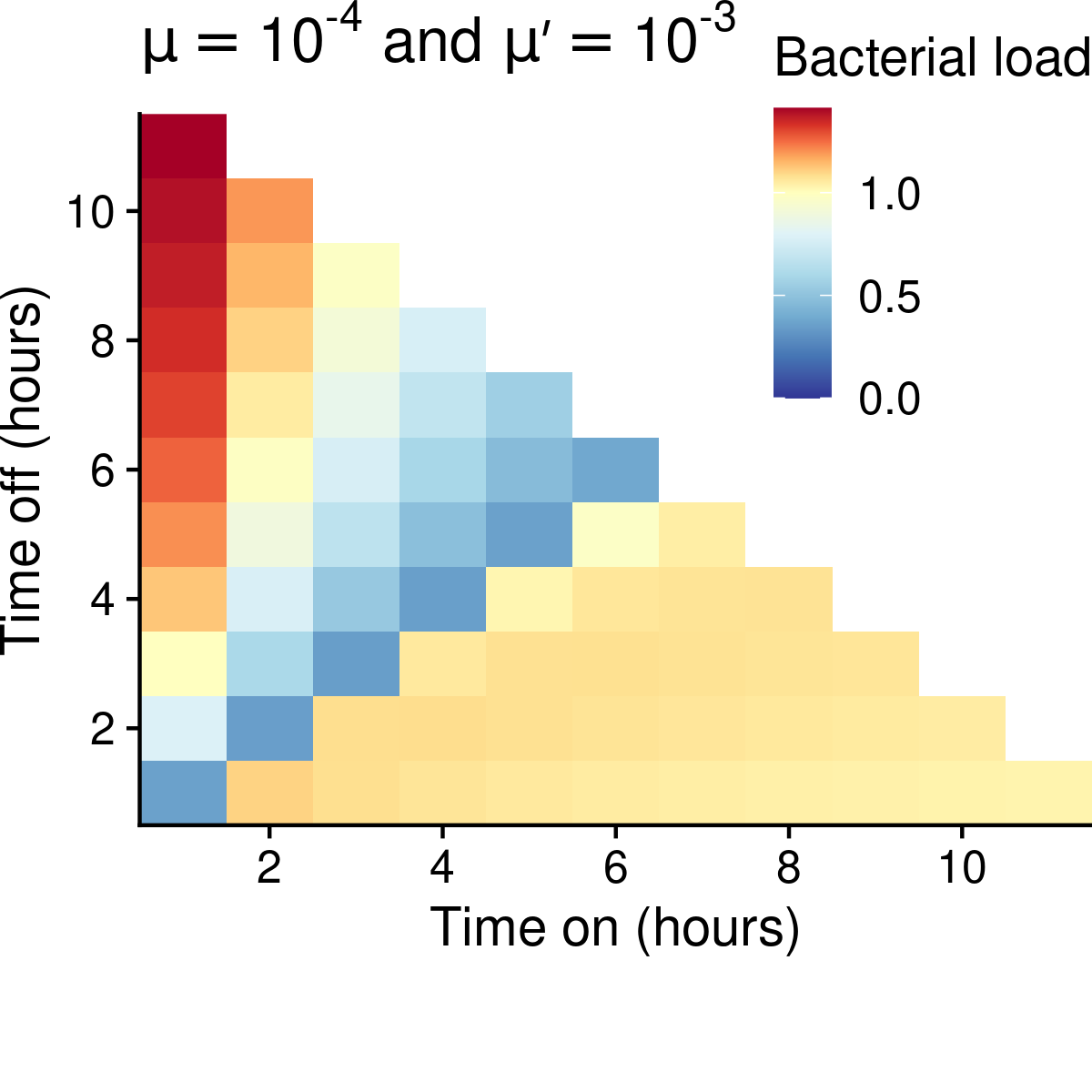}
    \end{subfigure} \\
    \begin{subfigure}[]{0.9\textwidth}
        \caption{}
        \includegraphics[width=\textwidth]{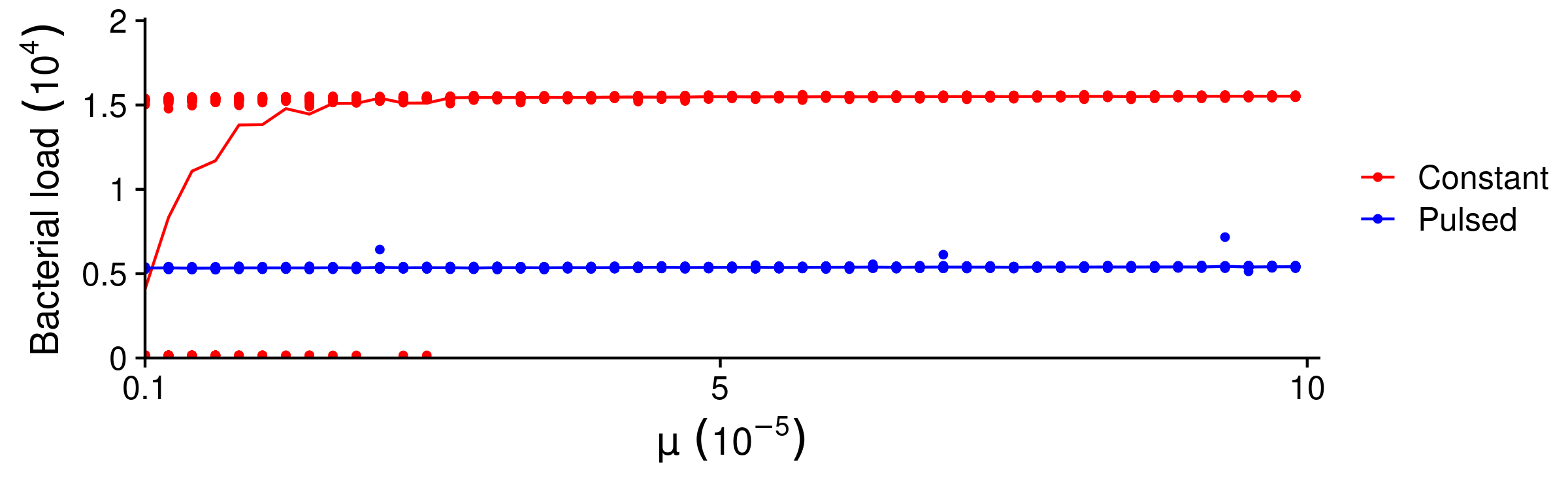} \label{fig:mu_g}
    \end{subfigure}
    \caption{Heatmaps of the average bacterial load over time from pulsed protocols relative to that of constant application of the antibiotic for $\mu = 10^{-6}, 10^{-5}, 10^{-4}$, $\mu' = \mu, 10\mu$, and $\kappa=4$. Protocols that matched the average outcome of the constant application therapy are coloured in yellow. Those protocols that did worse are in red, and those that did better are in blue. Panel \textbf{g} depicts the average bacterial load over time for constant and pulsed ($2$hrs on and off each) for various $\mu$. The points are the results for individual realizations and the curves their average.}
    \label{fig:mu}
\end{figure}

To explore how robust our results are to mutation rates, we considered various values of $\mu$ and $\mu'$. We can see the effects of various $\mu$ in the rows of Figure \ref{fig:mu}, which show that the pulsed protocols are more effective under a higher mutation rate. The first row depicts the case where there is a stress induced mutation rate from wild-type to mutant ($\mu'=10\mu$). The second row depicts the results where stress does not increase the mutation rate ($\mu'=\mu$). The stress induced mutation makes the antibiotic environment more conducive to generating resistance, and thus makes it harder to control the emergence of resistance. The impact is a small relative effect on the pulsed protocols vs.\ constant application protocols. Figure \ref{fig:mu_g} shows that the mutation rate impacts the constant application more so than the pulsed protocol. This result matches intuition; the higher the mutation rate, the less likely a constant application can eliminate the colony before a mutant arises and becomes established. In summary, the more evolvable the system, the better switching environments works.

\begin{figure}[!htb]
    \centering
    \begin{subfigure}[]{0.3\textwidth}
        \caption{}
        \includegraphics[width=\textwidth]{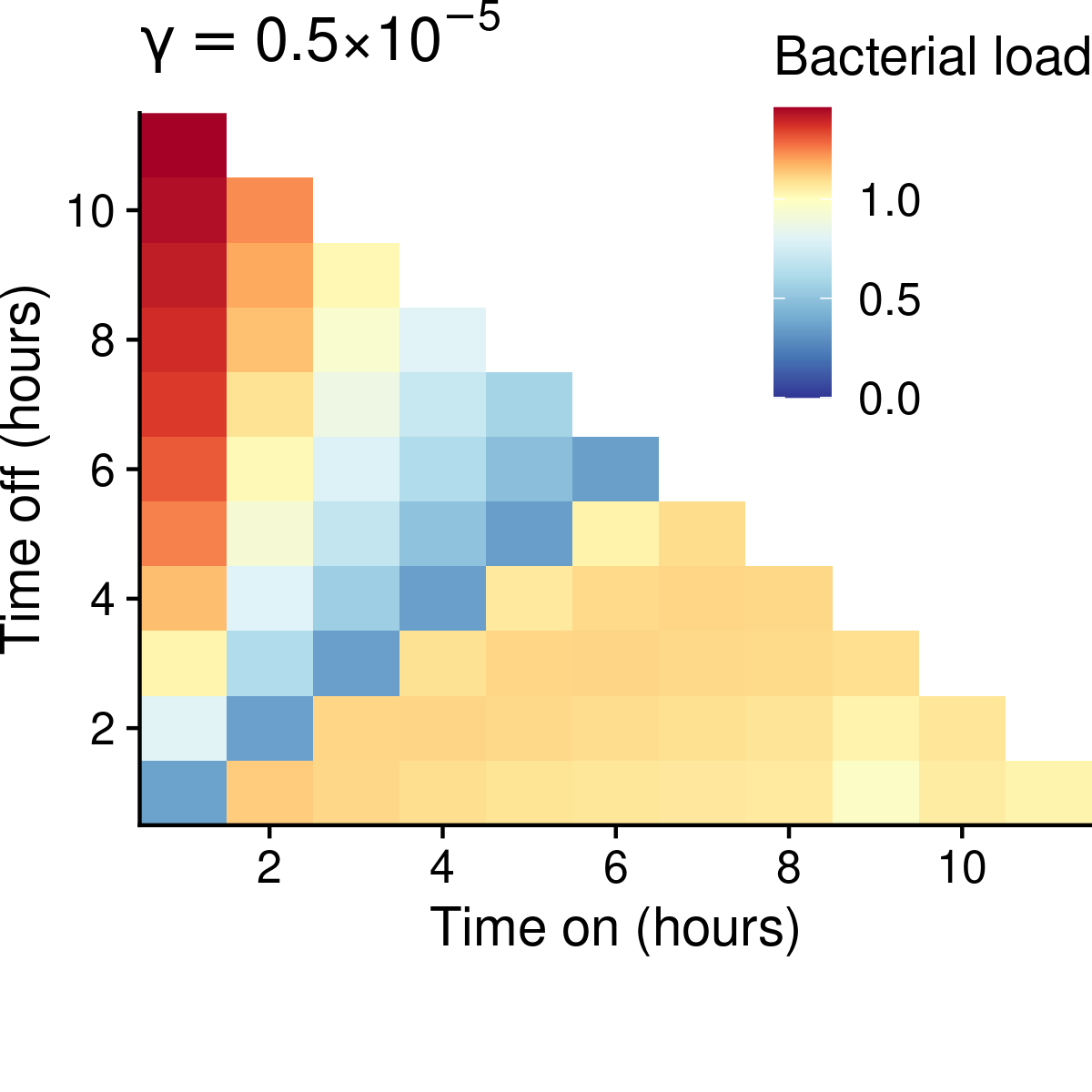}
    \end{subfigure}
    \begin{subfigure}[]{0.3\textwidth}
        \caption{}
        \includegraphics[width=\textwidth]{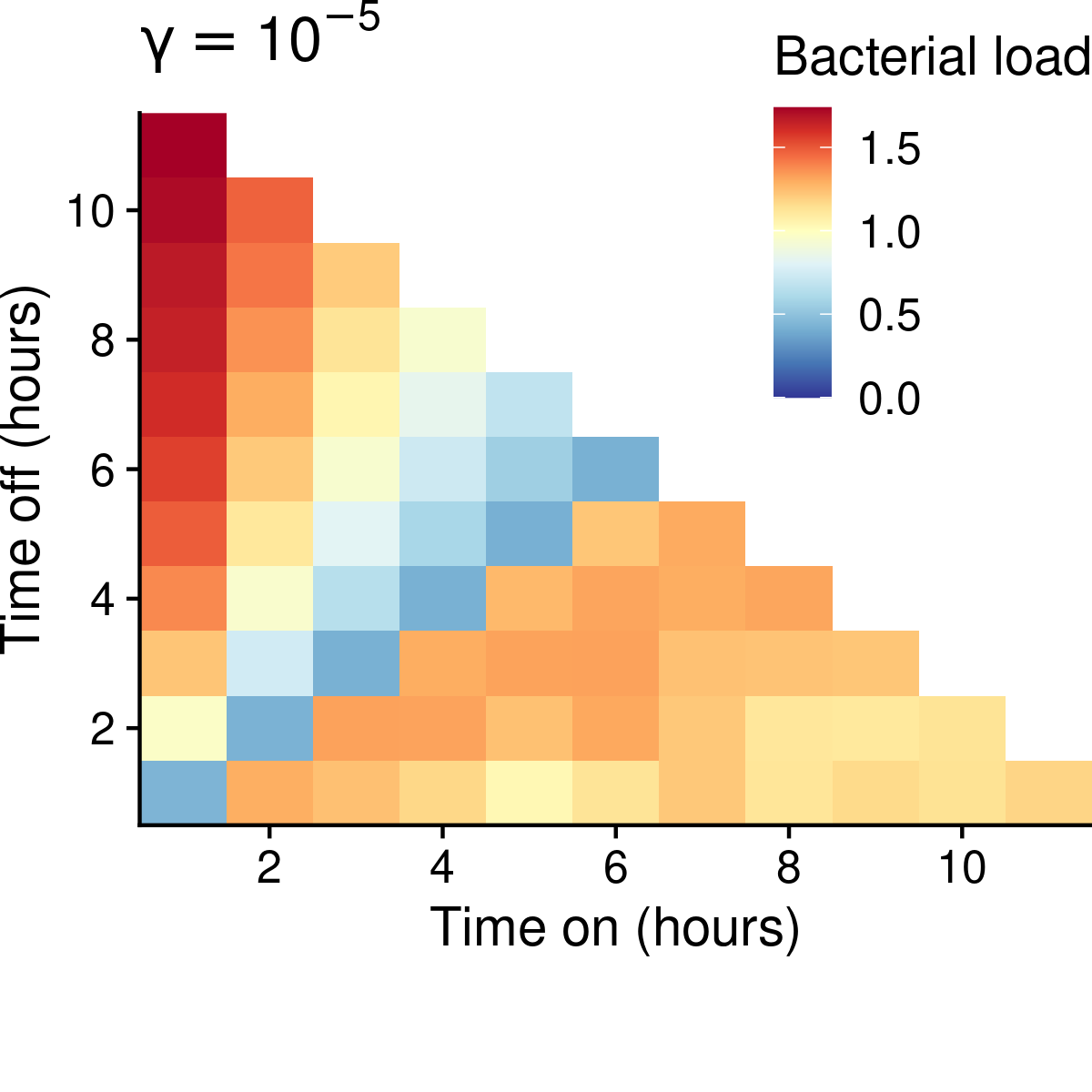}
    \end{subfigure}
    \begin{subfigure}[]{0.3\textwidth}
        \caption{}
        \includegraphics[width=\textwidth]{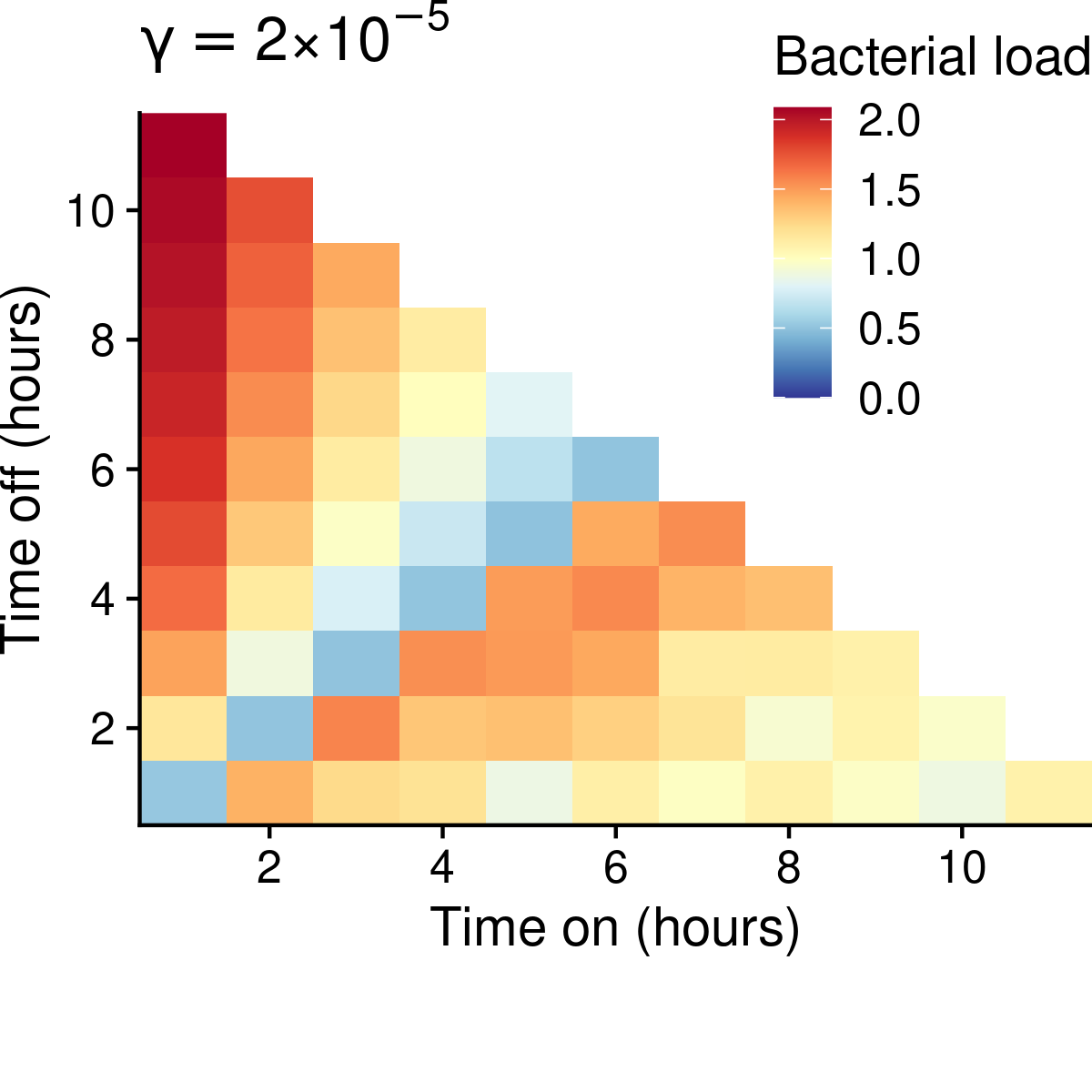}
    \end{subfigure} \\
    \begin{subfigure}[]{0.9\textwidth}
        \caption{}
        \includegraphics[width=\textwidth]{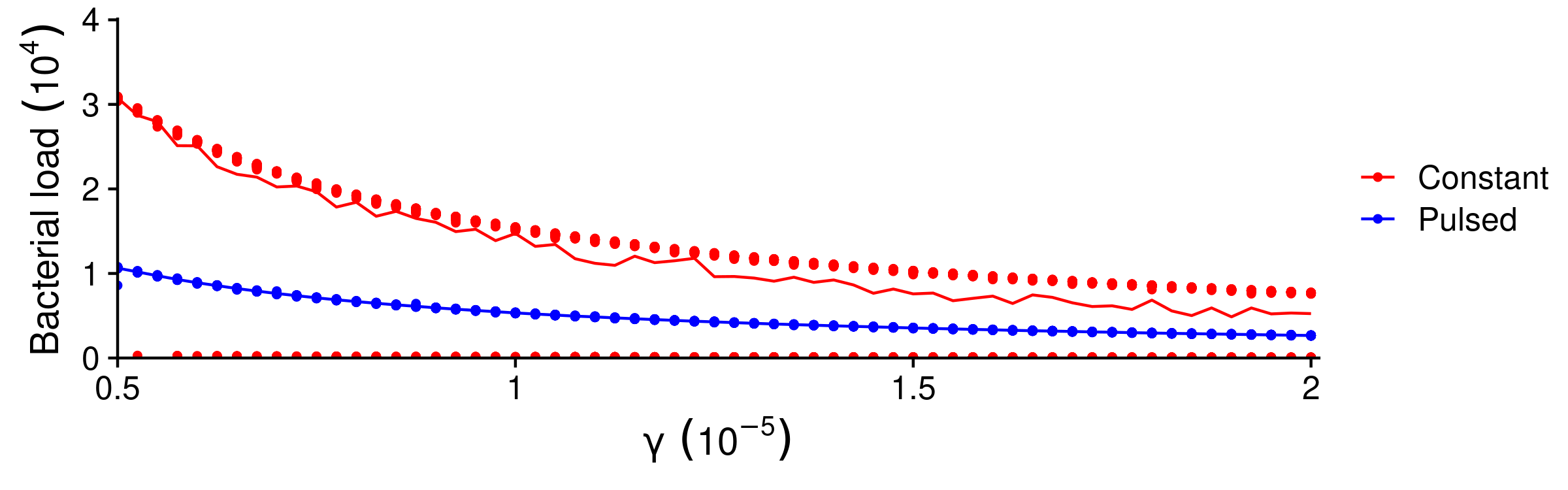} \label{fig:gamma_g}
    \end{subfigure}
    \caption{Heatmaps of the average bacterial load over time from pulsed protocols relative to that of constant application of the antibiotic for $\gamma = 0.5 \times 10^{-5}, 10^{-5}, 2 \times 10^{-5}$ and $\kappa=4$. Protocols that matched the average outcome of the constant application therapy are coloured in yellow. Those protocols that did worse are in red, and those that did better are in blue. Panel \textbf{d} depicts the average bacterial load over time for constant and pulsed ($2$hrs on and off each) for various $\gamma$. The points are the results for individual realizations and the curves their average.}
    \label{fig:gamma}
\end{figure}

In Figure \ref{fig:gamma} we explored the effect of varying the contact rate $\gamma$, and observed that switching was more effective for a low $\gamma$. For a high $\gamma$, the region where both types can grow is small, which magnifies the impact of stochastic effects leading to elimination of emerging mutants. Further, same-type competitive interactions are also more intense, and thus the population is driven to extinction more quickly. Figure \ref{fig:gamma_g} shows that the effectiveness of both constant application and pulsed protocols increases as $\gamma$ increases (and thus the carrying capacity is lower). However, the gap between the two shrinks. Thus, for a high $\gamma$ system, the average of the bacterial load of the individual realizations for the constant application is similarly to the low level of the pulsed protocol.

\section{Discussion}

Mathematical modelling has been important in the fight against resistance through increasing our understanding of the dynamics and emergence of resistance \cite{opatowski11}. This paper contributes to this endeavour by showing that the average bacterial load over time can be reduced and the emergence of resistant mutants can be mitigated via pulsed protocols. Previous research has shown that pulsed protocols with sufficiently long periods between switching can eliminate the population while rapidly changing environments are ineffective \cite{marrec20}. Our model would also exhibits this behaviour. Long durations between pulses can (re)establish the wild-type. Upon application of the antibiotic, the population would then either be eliminated or rescued by resistant mutants. However, in a clinical setting, this may not be feasible. Though quickly varying the environment can often not eliminate the population, it can suppress the bacterial load and resistance. This strategy can be thought of as "playing to not lose", which contrasts with a constant application (or long durations) where the bacteria are either eliminated or resistance flourishes (i.e.\ "playing to win or lose") \cite{fischer15}.

We found that pulsed protocols only work when competition is high. Else, at low population levels, resistance can be maintained. Previous empirical and theoretical research has also found the importance of high competition in containing an infection \cite{hansen20}. Further, pulsed protocols and competition can be effective in containing an infection even in a well-mixed population \cite{hansen20}. Spatial effects, such as those found in biofilms, could heighten the degree of competition and thus the effectiveness of the pulsed protocols. Since, spatial heterogeneity due to clumping could keep the competition level between different types high even when the population is small relative to the carrying capacity. The competition effects we consider can be interpreted crudely as arising from such effects.

Bactericides with significant post-antibiotic effects (PAE), such as fluoroquinolones, may hamper our control strategy. Such antimicrobials can impact the bacteria at sub-MIC levels long after they have been removed from the system \cite{shojaee00}, and as such can select for resistant bacteria after the antibiotic has been turned off. Additionally, sub MIC levels can lead to multidrug resistance through radical-induced mutagensis \cite{kohanski10}. Therefore, we must have a rapid dissipation of the antibiotic once below the MIC to prevent selection for the mutant (the range between the MIC and the point at which the susceptible strain is selected for) \cite{gullberg11}. PAE is frequently caused by antibiotics that impair DNA functioning. Hence, $\beta$-lactams, which inhibit cell wall production, are a good choice for our therapies. Antibiotics with a short half life would also be effective with our protocols.

Intermediate drug concentration is a dimension we did not consider here. The effects of concentration on the spread of the disease due to within host and between host dynamics have been found to be an important factor \cite{scire19}. It is not always beneficial to have a high concentration due to  a U-shaped probability of resistance emerging vs.\ drug concentration \cite{day16}. Here we only considered a specific concentration for all applications.

We restrict ourselves to the case where the system cannot be well monitored. Clearly, the best strategy is to alter the durations dependent on the state of the system, which can both maximize the duration of the antibiotic regime and prevent the emergence of resistance. The more repeated applications of the treatment, the less likely it will work. However, such observations may not be feasible, especially when the bacterial load is small and heterogeneous.

Future models could incorporate other biological factors. For example, the setting and source of resistance (source-sink dynamics) are important factors in controlling antibiotic resistance \cite{perron07}. The method by which resistance is spread is another important factor such as where plasmids confer resistance. In such a scenario, resistance can reemerge rapidly. Since, plasmids can remain in the population due to horizontal transfer even when the plasmid confers a cost \cite{lopatkin17}. Although, the transfer rate has been show to dramatically fall once the population is low \cite{handel15}. Compensatory mutations could also be added by which the cost to resistance could be reduced ($c$ or $\kappa$ could be reduced). However, we did explore the case where $c=0$, and found that though pulsed protocols were less effective than when there was a cost, they could still suppress the population and resistance (see Appendix \ref{app:furthersims} for these results). Future models could also incorporate more of the complexity of interactions between the bacteria and the patients' natural flora \cite{wade16, estrela18}.

More generally, our work fits within the theory of controlling evolving populations. Which, outside of bacteria, has been used to study cancer \cite{komarova06, katouli11, fischer15} and which families of chemotherapies will work best. Our model can be viewed as control of "species" in conflict under directed actions of an external force (in our case, the application of antibiotics) or under environmental fluctuation (which could be undirected). Though our aim here has been to lower the overall bacterial load, in some simulations (see Figure \ref{fig:tsPulse_fail_2}), alternating environments prevents one species dominating thereby sustaining coexistence. This temporal heterogeneity in competitiveness can thus act as a stabilizing mechanism that promotes diversity (as measured by the relative proportions of each type over time). This observation has broader theoretical implications to abundance and diversity of phenotypes or species competing with one another. In a switching environment, intermediate levels of interaction between different phenotypes or species can result in higher diversity. e.g.\, in our case, both phenotypes may coexist when the environment is varying while a constant environment leads to extinction or the resistant strain dominating (i.e\ less relative diversity than if both types coexist at low levels). Though our model is within a framework of bacterial competition, this phenomenon would apply to competitive Lotka-Volterra systems under environmental switching more generally. As such, we envisage further research that explores such phenomena under scenarios other than control of antibiotic resistance and through models related to our own.


\subsection*{Acknowledgements}

We would like to thank Plamen Kamenov and Giuseppe Forte for assistance with earlier versions of this project. This material is based upon work supported by the Defense Advanced Research Projects Agency under Contract No.\ HR0011-16-C-0062, and the University of Pennsylvania.

\subsection*{Author contributions}

Both authors contributed to the conception of the study and the final manuscript. B.M. developed the code for and analyzed the numerical simulations, and wrote the first draft.

\subsubsection*{Data availability}
The code to run the numerical simulations and make figures and the data for the figures are available at \url{https://github.com/bmorsky/antibioticresistance}.

\bibliography{main}

\begin{thebibliography}{10}

\bibitem{acar19}
{\sc Acar, N., and Cogan, N.~G.}
\newblock Enhanced disinfection of bacterial populations by nutrient and
  antibiotic challenge timing.
\newblock {\em Mathematical biosciences 313\/} (2019), 12--32.

\bibitem{shojaee00}
{\sc AliAbadi, F.~S., and Lees, P.}
\newblock Antibiotic treatment for animals: effect on bacterial population and
  dosage regimen optimisation.
\newblock {\em International Journal of Antimicrobial Agents 14}, 4 (2000),
  307--313.

\bibitem{andersson10}
{\sc Andersson, D.~I., and Hughes, D.}
\newblock Antibiotic resistance and its cost: is it possible to reverse
  resistance?
\newblock {\em Nature Reviews Microbiology 8}, 4 (2010), 260.

\bibitem{austin99}
{\sc Austin, D., and Anderson, R.}
\newblock Studies of antibiotic resistance within the patient, hospitals and
  the community using simple mathematical models.
\newblock {\em Philosophical Transactions of the Royal Society of London B:
  Biological Sciences 354}, 1384 (1999), 721--738.

\bibitem{baker18}
{\sc Baker, C.~M., Ferrari, M.~J., and Shea, K.}
\newblock Beyond dose: Pulsed antibiotic treatment schedules can maintain
  individual benefit while reducing resistance.
\newblock {\em Scientific Reports 8}, 1 (2018), 5866.

\bibitem{balaban04}
{\sc Balaban, N.~Q., Merrin, J., Chait, R., Kowalik, L., and Leibler, S.}
\newblock Bacterial persistence as a phenotypic switch.
\newblock {\em Science 305}, 5690 (2004), 1622--1625.

\bibitem{basra18}
{\sc Basra, P., Alsaadi, A., Bernal-Astrain, G., O’Sullivan, M.~L., Hazlett,
  B., Clarke, L.~M., Schoenrock, A., Pitre, S., and Wong, A.}
\newblock Fitness tradeoffs of antibiotic resistance in extraintestinal
  pathogenic escherichia coli.
\newblock {\em Genome Biology and Evolution 10}, 2 (2018), 667--679.

\bibitem{bhagunde11}
{\sc Bhagunde, P., Singh, R., Ledesma, K.~R., Chang, K.-T., Nikolaou, M., and
  Tam, V.~H.}
\newblock Modelling biphasic killing of fluoroquinolones: guiding optimal
  dosing regimen design.
\newblock {\em Journal of Antimicrobial Chemotherapy 66}, 5 (2011), 1079--1086.

\bibitem{bonhoeffer97}
{\sc Bonhoeffer, S., Lipsitch, M., and Levin, B.~R.}
\newblock Evaluating treatment protocols to prevent antibiotic resistance.
\newblock {\em Proceedings of the National Academy of Sciences 94}, 22 (1997),
  12106--12111.

\bibitem{borges03}
{\sc Borges-Walmsley, M.~I., McKeegan, K.~S., and Walmsley, A.~R.}
\newblock Structure and function of efflux pumps that confer resistance to
  drugs.
\newblock {\em Biochemical Journal 376}, 2 (2003), 313--338.

\bibitem{coates18}
{\sc Coates, J., Park, B.~R., Le, D., {\c{S}}im{\c{s}}ek, E., Chaudhry, W., and
  Kim, M.}
\newblock Antibiotic-induced population fluctuations and stochastic clearance
  of bacteria.
\newblock {\em eLife 7\/} (2018), e32976.

\bibitem{cogan06}
{\sc Cogan, N.}
\newblock Effects of persister formation on bacterial response to dosing.
\newblock {\em Journal of Theoretical Biology 238}, 3 (2006), 694--703.

\bibitem{cogan12}
{\sc Cogan, N., Brown, J., Darres, K., and Petty, K.}
\newblock Optimal control strategies for disinfection of bacterial populations
  with persister and susceptible dynamics.
\newblock {\em Antimicrobial Agents and Chemotherapy 56}, 9 (2012), 4816--4826.

\bibitem{czock07}
{\sc Czock, D., and Keller, F.}
\newblock Mechanism-based pharmacokinetic--pharmacodynamic modeling of
  antimicrobial drug effects.
\newblock {\em Journal of Pharmacokinetics and Pharmacodynamics 34}, 6 (2007),
  727--751.

\bibitem{day16}
{\sc Day, T., and Read, A.~F.}
\newblock Does high-dose antimicrobial chemotherapy prevent the evolution of
  resistance?
\newblock {\em PLoS Computational Biology 12}, 1 (2016), e1004689.

\bibitem{ender04}
{\sc Ender, M., McCallum, N., Adhikari, R., and Berger-B{\"a}chi, B.}
\newblock Fitness cost of sccmec and methicillin resistance levels in
  staphylococcus aureus.
\newblock {\em Antimicrobial Agents and Chemotherapy 48}, 6 (2004), 2295--2297.

\bibitem{estrela18}
{\sc Estrela, S., and Brown, S.~P.}
\newblock Community interactions and spatial structure shape selection on
  antibiotic resistant lineages.
\newblock {\em PLoS Computational Biology 14}, 6 (2018), e1006179.

\bibitem{fischer15}
{\sc Fischer, A., V{\'a}zquez-Garc{\'\i}a, I., and Mustonen, V.}
\newblock The value of monitoring to control evolving populations.
\newblock {\em Proceedings of the National Academy of Sciences 112}, 4 (2015),
  1007--1012.

\bibitem{gillespie76}
{\sc Gillespie, D.~T.}
\newblock A general method for numerically simulating the stochastic time
  evolution of coupled chemical reactions.
\newblock {\em Journal of Computational Physics 22}, 4 (1976), 403--434.

\bibitem{gloede09}
{\sc Gloede, J., Scheerans, C., Derendorf, H., and Kloft, C.}
\newblock In vitro pharmacodynamic models to determine the effect of
  antibacterial drugs.
\newblock {\em Journal of Antimicrobial Chemotherapy 65}, 2 (2009), 186--201.

\bibitem{gonze18}
{\sc Gonze, D., Coyte, K.~Z., Lahti, L., and Faust, K.}
\newblock Microbial communities as dynamical systems.
\newblock {\em Current opinion in microbiology 44\/} (2018), 41--49.

\bibitem{greulich17}
{\sc Greulich, P., Dole{\v{z}}al, J., Scott, M., Evans, M.~R., and Allen,
  R.~J.}
\newblock Predicting the dynamics of bacterial growth inhibition by
  ribosome-targeting antibiotics.
\newblock {\em Physical biology 14}, 6 (2017), 065005.

\bibitem{gullberg11}
{\sc Gullberg, E., Cao, S., Berg, O.~G., Ilb{\"a}ck, C., Sandegren, L., Hughes,
  D., and Andersson, D.~I.}
\newblock Selection of resistant bacteria at very low antibiotic
  concentrations.
\newblock {\em PLoS Pathogens 7}, 7 (2011), e1002158.

\bibitem{handel15}
{\sc H{\"a}ndel, N., Otte, S., Jonker, M., Brul, S., and ter Kuile, B.~H.}
\newblock Factors that affect transfer of the inci1 $\beta$-lactam resistance
  plasmid pesbl-283 between e. coli strains.
\newblock {\em PloS One 10}, 4 (2015), e0123039.

\bibitem{hansen20}
{\sc Hansen, E., Karslake, J., Woods, R.~J., Read, A.~F., and Wood, K.~B.}
\newblock Antibiotics can be used to contain drug-resistant bacteria by
  maintaining sufficiently large sensitive populations.
\newblock {\em PLoS Biology 18}, 5 (2020), e3000713.

\bibitem{hauert08}
{\sc Hauert, C., Wakano, J.~Y., and Doebeli, M.}
\newblock Ecological public goods games: cooperation and bifurcation.
\newblock {\em Theoretical Population Biology 73}, 2 (2008), 257--263.

\bibitem{huang15}
{\sc Huang, W., Hauert, C., and Traulsen, A.}
\newblock Stochastic game dynamics under demographic fluctuations.
\newblock {\em Proceedings of the National Academy of Sciences 112}, 29 (2015),
  9064--9069.

\bibitem{katouli11}
{\sc Katouli, A.~A., and Komarova, N.~L.}
\newblock The worst drug rule revisited: mathematical modeling of cyclic cancer
  treatments.
\newblock {\em Bulletin of Mathematical Biology 73}, 3 (2011), 549--584.

\bibitem{kohanski10}
{\sc Kohanski, M.~A., DePristo, M.~A., and Collins, J.~J.}
\newblock Sublethal antibiotic treatment leads to multidrug resistance via
  radical-induced mutagenesis.
\newblock {\em Molecular Cell 37}, 3 (2010), 311--320.

\bibitem{komarova06}
{\sc Komarova, N.}
\newblock Stochastic modeling of drug resistance in cancer.
\newblock {\em Journal of Theoretical Biology 239}, 3 (2006), 351--366.

\bibitem{kuban04}
{\sc Kuban, W., Jonczyk, P., Gawel, D., Malanowska, K., Schaaper, R.~M., and
  Fijalkowska, I.~J.}
\newblock Role of escherichia coli dna polymerase iv in in vivo replication
  fidelity.
\newblock {\em Journal of Bacteriology 186}, 14 (2004), 4802--4807.

\bibitem{kussell05}
{\sc Kussell, E., Kishony, R., Balaban, N.~Q., and Leibler, S.}
\newblock Bacterial persistence: a model of survival in changing environments.
\newblock {\em Genetics 169}, 4 (2005), 1807--1814.

\bibitem{laxminarayan13}
{\sc Laxminarayan, R., Duse, A., Wattal, C., Zaidi, A.~K., Wertheim, H.~F.,
  Sumpradit, N., Vlieghe, E., Hara, G.~L., Gould, I.~M., Goossens, H., et~al.}
\newblock Antibiotic resistance—the need for global solutions.
\newblock {\em The Lancet Infectious Diseases 13}, 12 (2013), 1057--1098.

\bibitem{lipsitch97}
{\sc Lipsitch, M., and Levin, B.~R.}
\newblock The population dynamics of antimicrobial chemotherapy.
\newblock {\em Antimicrobial Agents and Chemotherapy 41}, 2 (1997), 363--373.

\bibitem{lopatkin17}
{\sc Lopatkin, A.~J., Meredith, H.~R., Srimani, J.~K., Pfeiffer, C., Durrett,
  R., and You, L.}
\newblock Persistence and reversal of plasmid-mediated antibiotic resistance.
\newblock {\em Nature Communications 8}, 1 (2017), 1689.

\bibitem{marrec20}
{\sc Marrec, L., and Bitbol, A.-F.}
\newblock Resist or perish: Fate of a microbial population subjected to a
  periodic presence of antimicrobial.
\newblock {\em PLoS computational biology 16}, 4 (2020), e1007798.

\bibitem{martinez02}
{\sc Mart{\'\i}nez, J.~L., and Baquero, F.}
\newblock Interactions among strategies associated with bacterial infection:
  pathogenicity, epidemicity, and antibiotic resistance.
\newblock {\em Clinical Microbiology Reviews 15}, 4 (2002), 647--679.

\bibitem{melnyk15}
{\sc Melnyk, A.~H., Wong, A., and Kassen, R.}
\newblock The fitness costs of antibiotic resistance mutations.
\newblock {\em Evolutionary Applications 8}, 3 (2015), 273--283.

\bibitem{nielsen11}
{\sc Nielsen, E.~I., Cars, O., and Friberg, L.~E.}
\newblock Pk/pd indices of antibiotics predicted by a semi-mechanistic pkpd
  model--a step towards model-based dose optimization.
\newblock {\em Antimicrobial Agents and Chemotherapy\/} (2011), AAC--00182.

\bibitem{nielsen13}
{\sc Nielsen, E.~I., and Friberg, L.~E.}
\newblock Pharmacokinetic-pharmacodynamic modeling of antibacterial drugs.
\newblock {\em Pharmacological Reviews 65}, 3 (2013), 1053--1090.

\bibitem{novozhilov06}
{\sc Novozhilov, A.~S., Karev, G.~P., and Koonin, E.~V.}
\newblock Biological applications of the theory of birth-and-death processes.
\newblock {\em Briefings in Bioinformatics 7}, 1 (2006), 70--85.

\bibitem{opatowski11}
{\sc Opatowski, L., Guillemot, D., Boelle, P.-Y., and Temime, L.}
\newblock Contribution of mathematical modeling to the fight against bacterial
  antibiotic resistance.
\newblock {\em Current Opinion in Infectious Diseases 24}, 3 (2011), 279--287.

\bibitem{pacheco17}
{\sc Pacheco, J.~O., Alvarez-Ortega, C., Rico, M.~A., and Mart{\'\i}nez, J.~L.}
\newblock Metabolic compensation of fitness costs is a general outcome for
  antibiotic-resistant pseudomonas aeruginosa mutants overexpressing efflux
  pumps.
\newblock {\em mBio 8}, 4 (2017), e00500--17.

\bibitem{perfeito07}
{\sc Perfeito, L., Fernandes, L., Mota, C., and Gordo, I.}
\newblock Adaptive mutations in bacteria: high rate and small effects.
\newblock {\em Science 317}, 5839 (2007), 813--815.

\bibitem{perron07}
{\sc Perron, G.~G., Gonzalez, A., and Buckling, A.}
\newblock Source--sink dynamics shape the evolution of antibiotic resistance
  and its pleiotropic fitness cost.
\newblock {\em Proceedings of the Royal Society of London B: Biological
  Sciences 274}, 1623 (2007), 2351--2356.

\bibitem{poole02}
{\sc Poole, K.}
\newblock Mechanisms of bacterial biocide and antibiotic resistance.
\newblock {\em Journal of Applied Microbiology 92\/} (2002), 55S--64S.

\bibitem{rackauckas17}
{\sc Rackauckas, C., and Nie, Q.}
\newblock Differentialequations. jl – a performant and feature-rich ecosystem
  for solving differential equations in julia.
\newblock {\em Journal of Open Research Software 5}, 1 (2017).
\newblock DOI:http://doi.org/10.5334/jors.151.

\bibitem{schmidt09}
{\sc Schmidt, S., Sabarinath, S.~N., Barbour, A., Abbanat, D., Manitpisitkul,
  P., Sha, S., and Derendorf, H.}
\newblock Pharmacokinetic-pharmacodynamic modeling of the in vitro activities
  of oxazolidinone antimicrobial agents against methicillin-resistant
  staphylococcus aureus.
\newblock {\em Antimicrobial Agents and Chemotherapy 53}, 12 (2009),
  5039--5045.

\bibitem{scire19}
{\sc Scire, J., Hoz{\'e}, N., and Uecker, H.}
\newblock Aggressive or moderate drug therapy for infectious diseases?
  trade-offs between different treatment goals at the individual and population
  levels.
\newblock {\em PLoS Computational Biology 15}, 8 (2019), e1007223.

\bibitem{sharma15}
{\sc Sharma, B., Brown, A.~V., Matluck, N.~E., Hu, L.~T., and Lewis, K.}
\newblock Borrelia burgdorferi, the causative agent of lyme disease, forms
  drug-tolerant persister cells.
\newblock {\em Antimicrobial Agents and Chemotherapy 59}, 8 (2015), 4616--4624.

\bibitem{stein13}
{\sc Stein, R.~R., Bucci, V., Toussaint, N.~C., Buffie, C.~G., R{\"a}tsch, G.,
  Pamer, E.~G., Sander, C., and Xavier, J.~B.}
\newblock Ecological modeling from time-series inference: insight into dynamics
  and stability of intestinal microbiota.
\newblock {\em PLoS Comput Biol 9}, 12 (2013), e1003388.

\bibitem{sun14}
{\sc Sun, J., Deng, Z., and Yan, A.}
\newblock Bacterial multidrug efflux pumps: mechanisms, physiology and
  pharmacological exploitations.
\newblock {\em Biochemical and Biophysical Research Communications 453}, 2
  (2014), 254--267.

\bibitem{tam08}
{\sc Tam, V.~H., Ledesma, K.~R., Vo, G., Kabbara, S., Lim, T.-P., and Nikolaou,
  M.}
\newblock Pharmacodynamic modeling of aminoglycosides against pseudomonas
  aeruginosa and acinetobacter baumannii: identifying dosing regimens to
  suppress resistance development.
\newblock {\em Antimicrobial Agents and Chemotherapy 52}, 11 (2008),
  3987--3993.

\bibitem{tam05}
{\sc Tam, V.~H., Louie, A., Deziel, M.~R., Liu, W., Leary, R., and Drusano,
  G.~L.}
\newblock Bacterial-population responses to drug-selective pressure:
  examination of garenoxacin’s effect on pseudomonas aeruginosa.
\newblock {\em The Journal of Infectious Diseases 192}, 3 (2005), 420--428.

\bibitem{tam07}
{\sc Tam, V.~H., Schilling, A.~N., Poole, K., and Nikolaou, M.}
\newblock Mathematical modelling response of pseudomonas aeruginosa to
  meropenem.
\newblock {\em Journal of Antimicrobial Chemotherapy 60}, 6 (2007), 1302--1309.

\bibitem{vanBambeke00}
{\sc Van~Bambeke, F., Balzi, E., and Tulkens, P.~M.}
\newblock Antibiotic efflux pumps.
\newblock {\em Biochemical Pharmacology 60}, 4 (2000), 457--470.

\bibitem{wade16}
{\sc Wade, M.~J., Harmand, J., Benyahia, B., Bouchez, T., Chaillou, S., Cloez,
  B., Godon, J.-J., Boudjemaa, B.~M., Rapaport, A., Sari, T., et~al.}
\newblock Perspectives in mathematical modelling for microbial ecology.
\newblock {\em Ecological Modelling 321\/} (2016), 64--74.

\bibitem{wang17}
{\sc Wang-Kan, X., Blair, J.~M., Chirullo, B., Betts, J., La~Ragione, R.~M.,
  Ivens, A., Ricci, V., Opperman, T.~J., and Piddock, L.~J.}
\newblock Lack of acrb efflux function confers loss of virulence on salmonella
  enterica serovar typhimurium.
\newblock {\em mBio 8}, 4 (2017), e00968--17.

\bibitem{webber03}
{\sc Webber, M., and Piddock, L.}
\newblock The importance of efflux pumps in bacterial antibiotic resistance.
\newblock {\em Journal of Antimicrobial Chemotherapy 51}, 1 (2003), 9--11.

\bibitem{wilkinson11}
{\sc Wilkinson, D.~J.}
\newblock {\em Stochastic modelling for systems biology}.
\newblock CRC press, 2011.

\bibitem{zhang12}
{\sc Zhang, Y., Yew, W.~W., and Barer, M.~R.}
\newblock Targeting persisters for tuberculosis control.
\newblock {\em Antimicrobial Agents and Chemotherapy 56}, 5 (2012), 2223--2230.

\end{thebibliography}
\bibliographystyle{acm}

\appendix

\section{Derivation of the Fokker-Planck and mean field equations} \label{app:derivation}

Here we derive the mean field equations. This model includes a stress induced mutation rate $\mu'$ (see Appendix \label{app:stress} for details). Let $P(X,Y,t)$ be the probability of $X$ and $Y$ numbers of each type at time $t$. Then, given that we have a two dimensional Markov process in continuous time, the master equation is

\begin{align}
\frac{\partial P(X,Y,t)}{\partial t} &= b(X-1)P(X-1,Y,t) + (b-c)(Y-1)P(X,Y-1,t) \notag \\
&+ \Big[d + \alpha\bar{A} + \gamma(X+1) + \frac{\gamma}{\kappa}Y\Big](X+1)P(X+1,Y,t) \notag \\
&+ \Big[d + \alpha'\bar{A} + \gamma\kappa X + \gamma(Y+1)\Big](Y+1)P(X,Y+1,t) \notag \\
&+ \mu(Y+1)P(X-1,Y+1,t) + (\mu(1-\bar{A})+\mu'\bar{A})(X+1)P(X+1,Y-1,t) \notag \\
&- \Big[(b+d+(1-\bar{A})\mu+(\alpha+\mu')\bar{A})X + (b-c+d+\mu+\alpha'\bar{A})Y \notag \\
&+ \gamma X^2 + \Big(\kappa+\frac{1}{\kappa}\Big)\gamma XY + \gamma Y^2\Big]P(X,Y,t).
\end{align}
\noindent Taking the time derivative of the average values $\m{X(t)} = \sum_{X,Y}P(X,Y,t)$ and $\m{Y(t)} = \sum_{X,Y}P(X,Y,t)$, we can find the mean field model:
\begin{align}
\dot{\m{X}} &= (b - d - \mu(1-\bar{A}) - (\alpha+\mu')\bar{A})\m{X} + \mu\m{Y} - \gamma\m{X}^2 - \frac{\gamma}{\kappa}\m{X}\m{Y}, \\
\dot{\m{Y}} &= (b - c - d - \mu - \alpha'\bar{A})\m{Y} + (\mu(1-\bar{A}) + \mu'\bar{A})\m{X} - \gamma\kappa\m{X}\m{Y} - \gamma\m{Y}^2.
\end{align}
\noindent Here we are assuming that $\m{XX} = \m{X}^2$, $\m{XY} = \m{X}\m{Y}$, and $\m{Y^2} =\m{Y}^2$.

\section{Further simulations results} \label{app:furthersims}

\begin{figure}[!htb]
\centering
    \includegraphics[width=\textwidth]{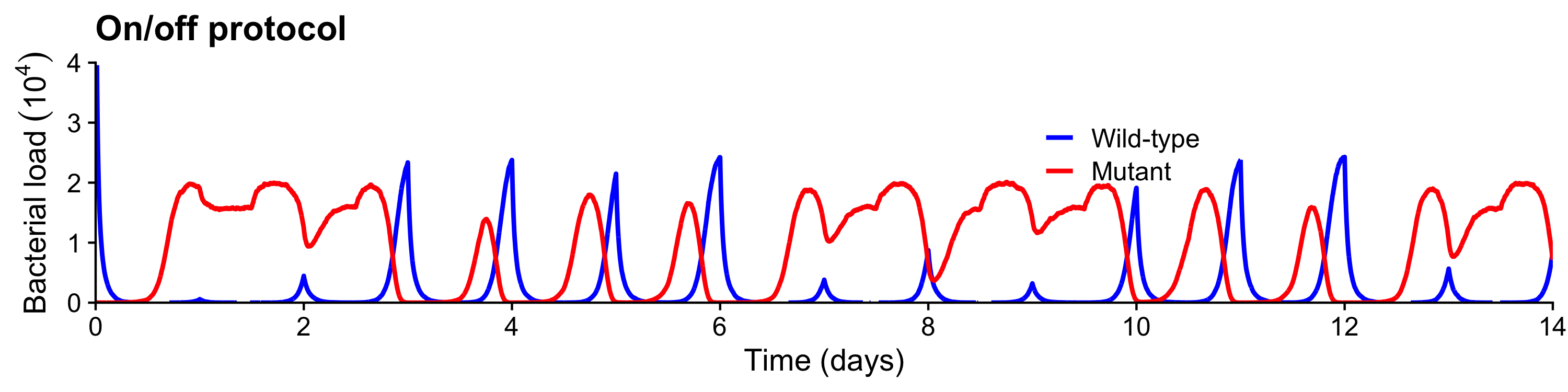}
\caption{Representative times series for a switching protocol (on and off for $12$hrs each). $\kappa=4$ and the remaining parameters are from Table \ref{param}.}
\label{fig:LVts_24hr}
\end{figure}

Here we depicts other simulations results. In Figure \ref{fig:LVts_24hr} we depict an example time series for a one day on and off protocol where the bacteria is not eliminated. Longer durations on or off can lead to the population reaching the carrying capacity, after which they will be markedly less successful at suppressing the bacteria than the constant application therapy.

\begin{figure}[!htb]
    \centering
    \begin{subfigure}[]{0.3\textwidth}
        \caption{}
        \includegraphics[width=\textwidth]{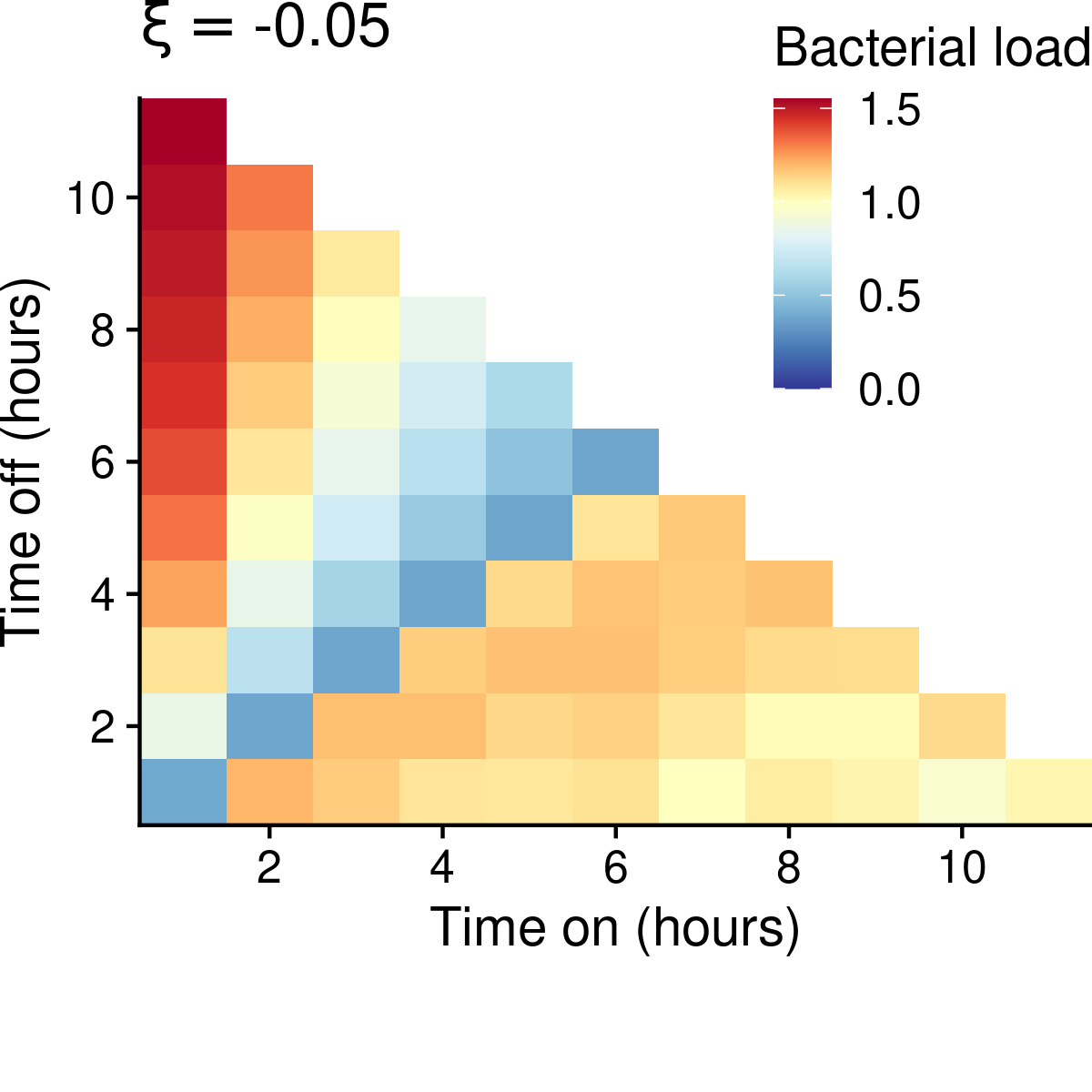}
    \end{subfigure}
    \begin{subfigure}[]{0.3\textwidth}
        \caption{}
        \includegraphics[width=\textwidth]{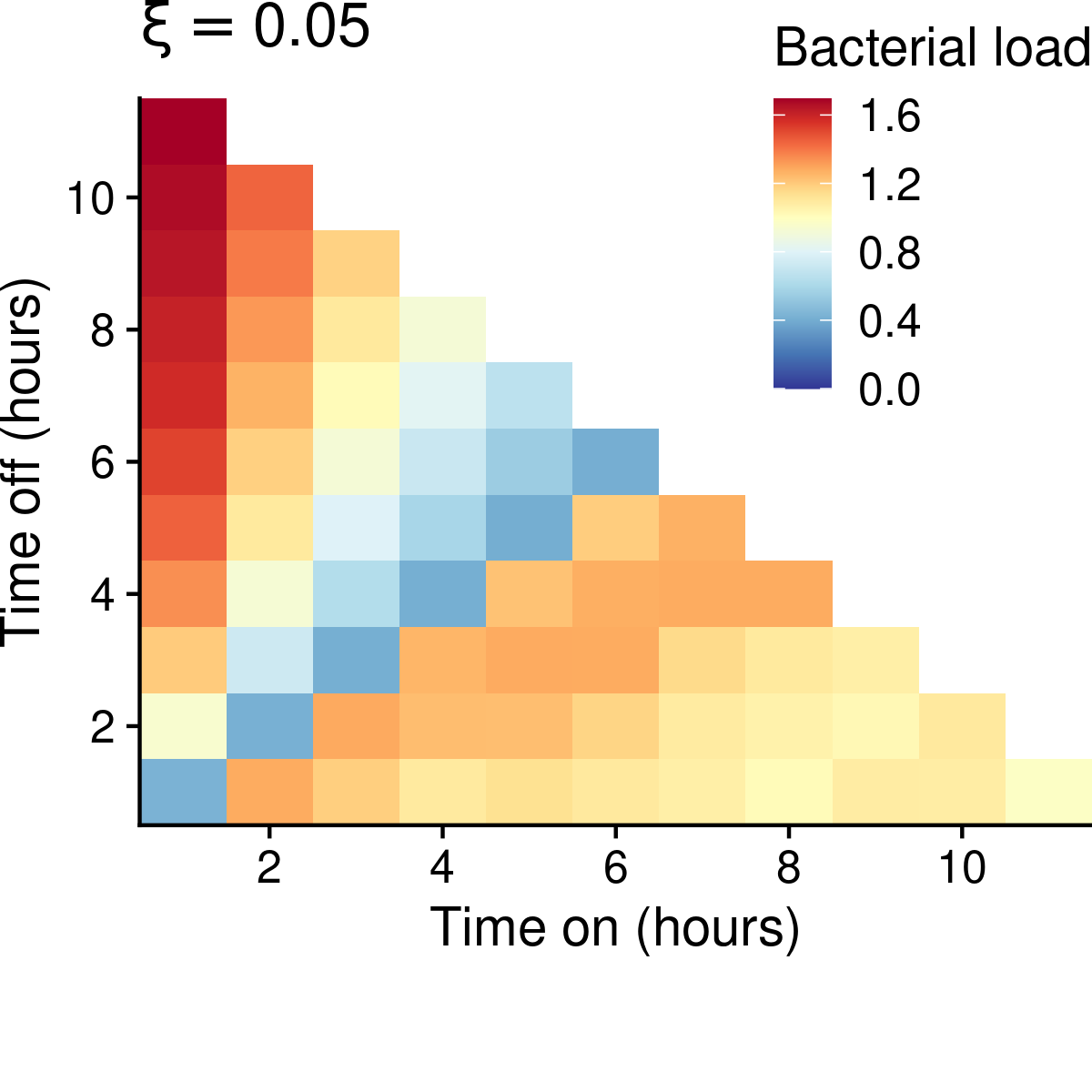}
    \end{subfigure}
        \begin{subfigure}[]{0.3\textwidth}
        \caption{}
        \includegraphics[width=\textwidth]{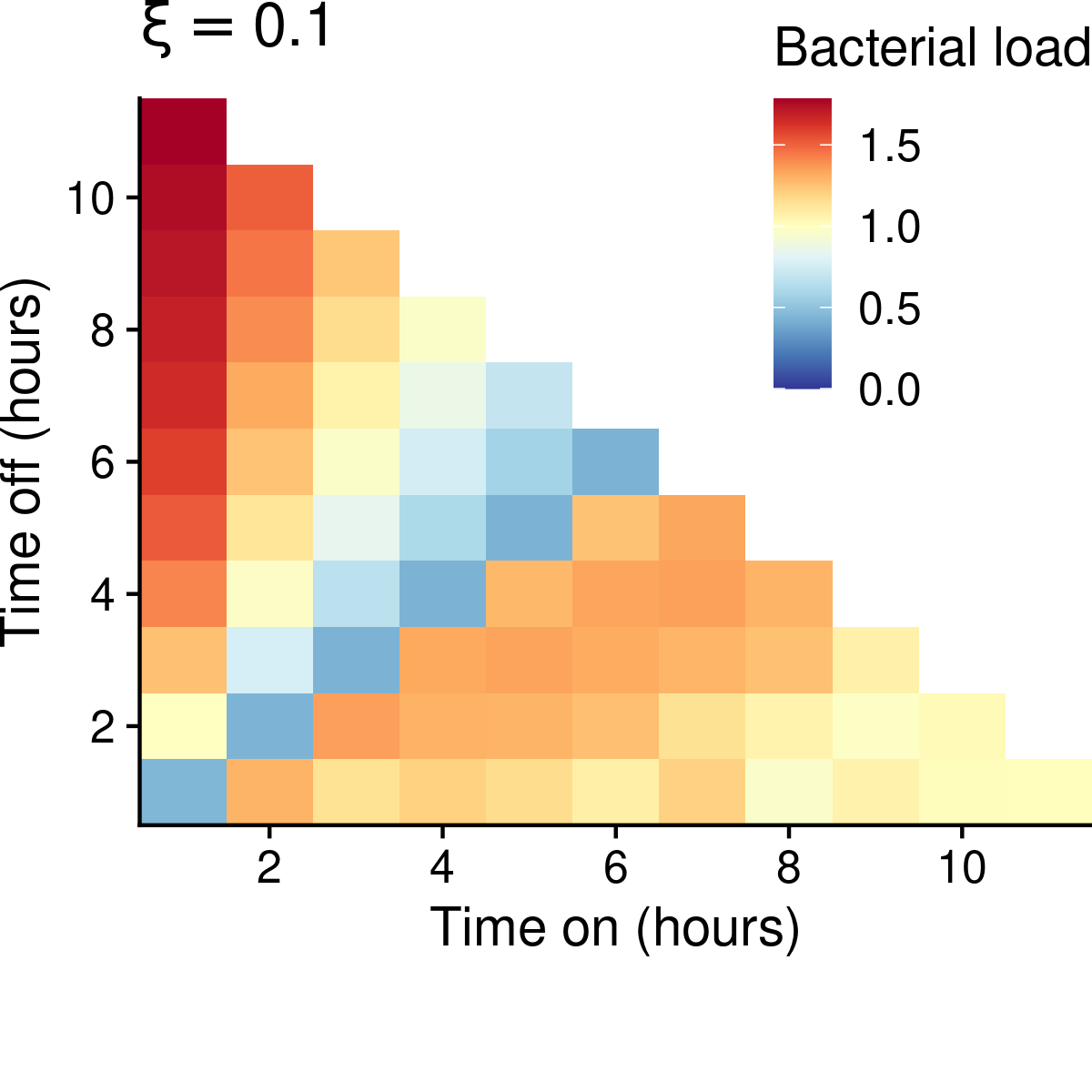}
    \end{subfigure} \\
        \begin{subfigure}[]{0.3\textwidth}
        \caption{}
        \includegraphics[width=\textwidth]{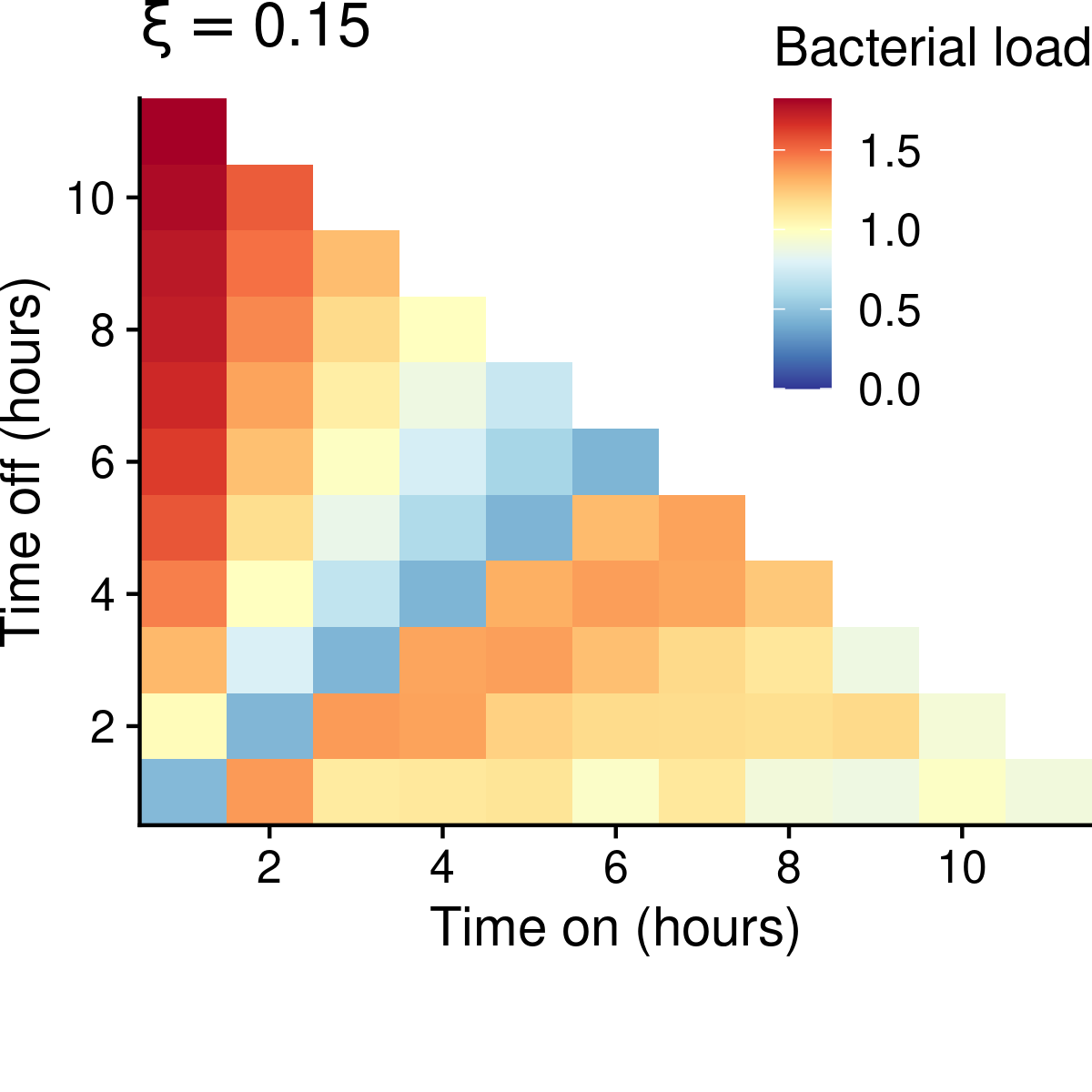}
    \end{subfigure}
        \begin{subfigure}[]{0.3\textwidth}
        \caption{}
        \includegraphics[width=\textwidth]{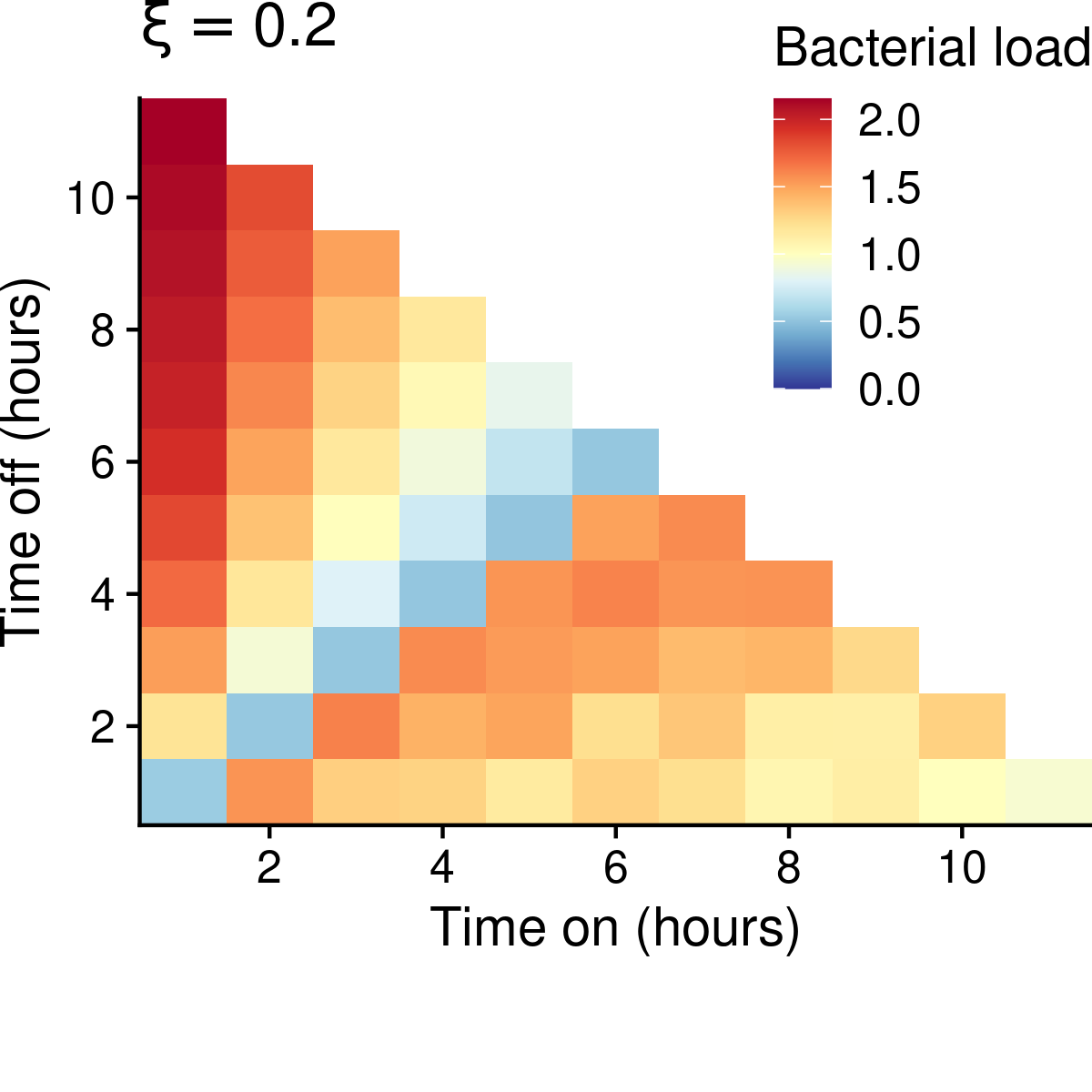}
    \end{subfigure}
    \begin{subfigure}[]{0.3\textwidth}
        \caption{}
        \includegraphics[width=\textwidth]{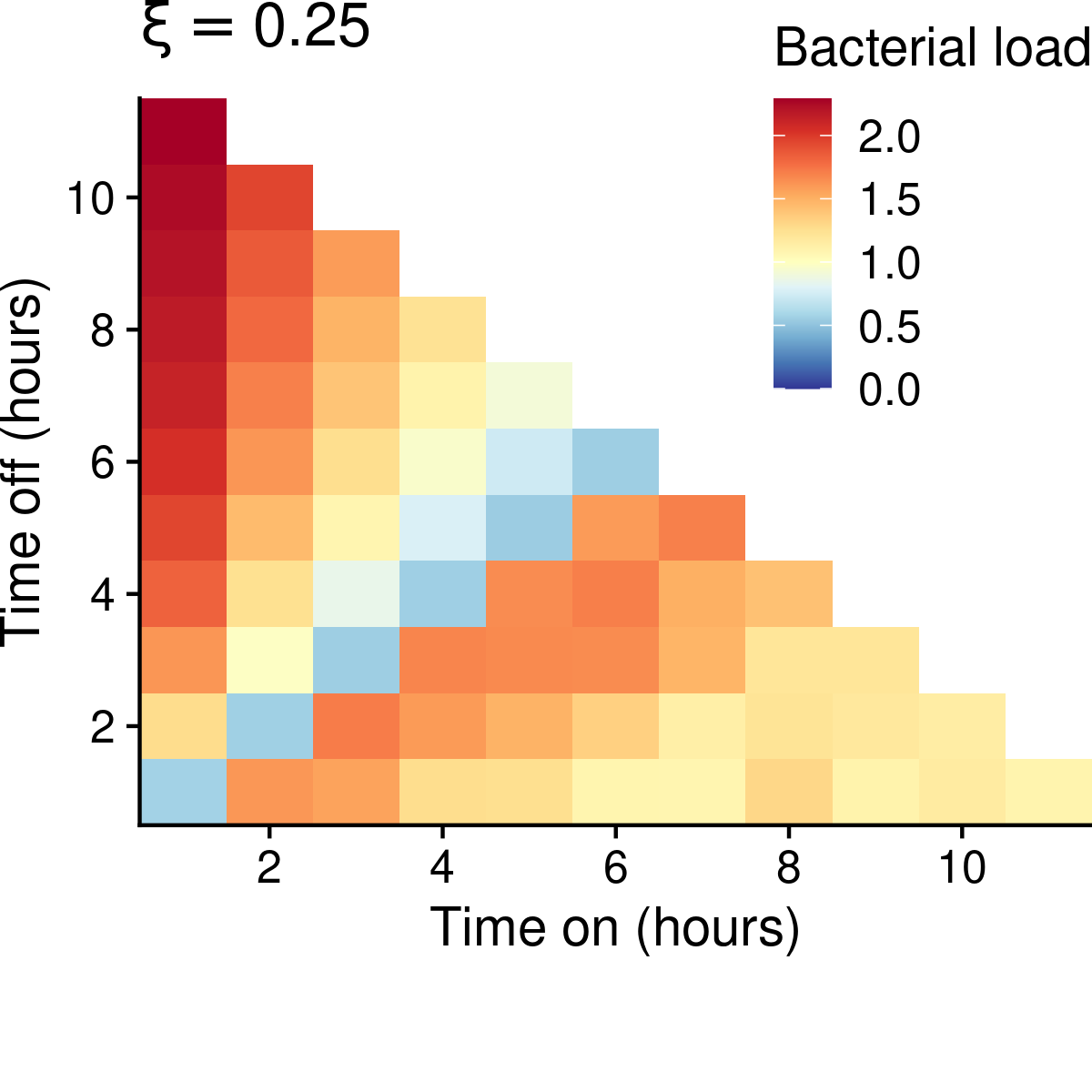}
    \end{subfigure} \\
    \begin{subfigure}[]{0.9\textwidth}
        \caption{}
        \includegraphics[width=\textwidth]{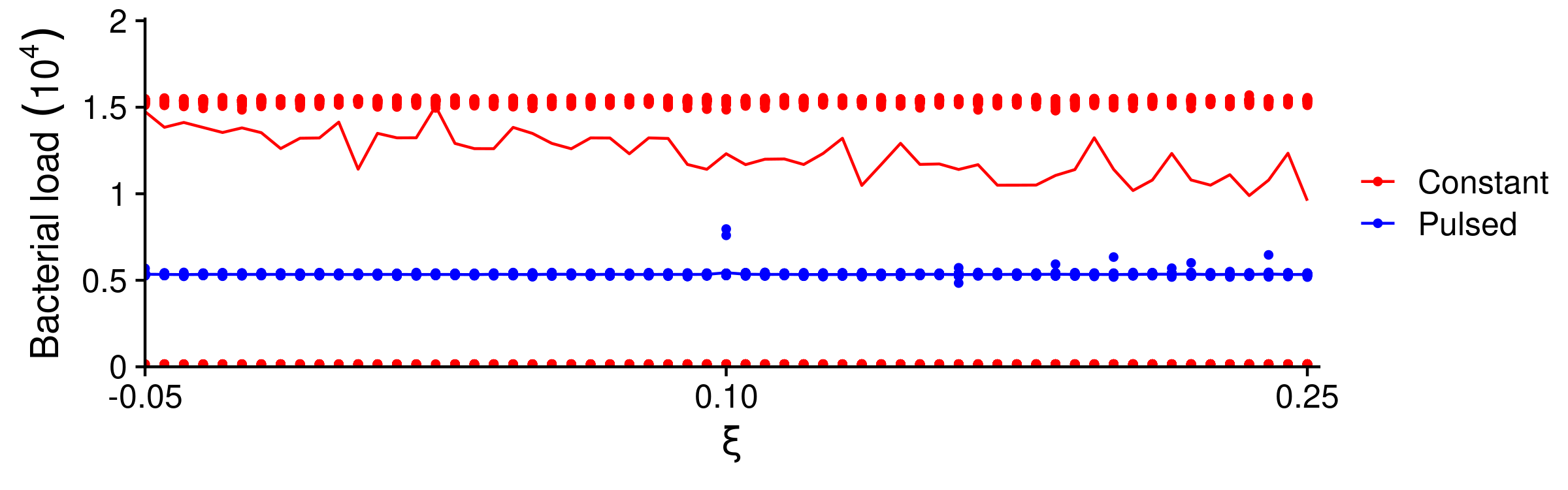} \label{fig:highstochasticity_g}
    \end{subfigure}
    \caption{Heatmaps of the average bacterial load over time from pulsed protocols relative to that of constant application of the antibiotic for $\xi = -0.05, 0.05, 0.1, 0.15, 0.2, 0.25$ and $\kappa=4$. Protocols that matched the average outcome of the constant application are coloured in yellow, those worse in red, and those better in blue. Panel \textbf{g} depicts the average bacterial load over time for constant and pulsed ($2$hrs on and off each) for various $\xi$ The points are the results for individual realizations and the curves the average.}
    \label{fig:highstochasticity}
\end{figure}

In order to explore the effects of greater stochasticity in the birth and death rates on the outcomes of these therapies, we ran simulations with birth and death rates $b+\xi$ and $d+\xi$, which keeps the mean growth rates unchanged. Figure \ref{fig:highstochasticity} depicts the results. We see that if the higher stochasticity can reduce the relative effectiveness of pulsed protocols. However, this effect is not large. In Figure \ref{fig:highstochasticity_g}, we can see that the gulf between the outcomes of constant applications and pulsed protocols decreases as $\xi$ increases.

\begin{figure}[!htbp]
        \centering
        \begin{subfigure}[]{0.9\textwidth}
            \caption{}
            \includegraphics[width=\textwidth]{img/alpha.png}
        \end{subfigure}
        \begin{subfigure}[]{0.9\textwidth}
            \caption{}
            \includegraphics[width=\textwidth]{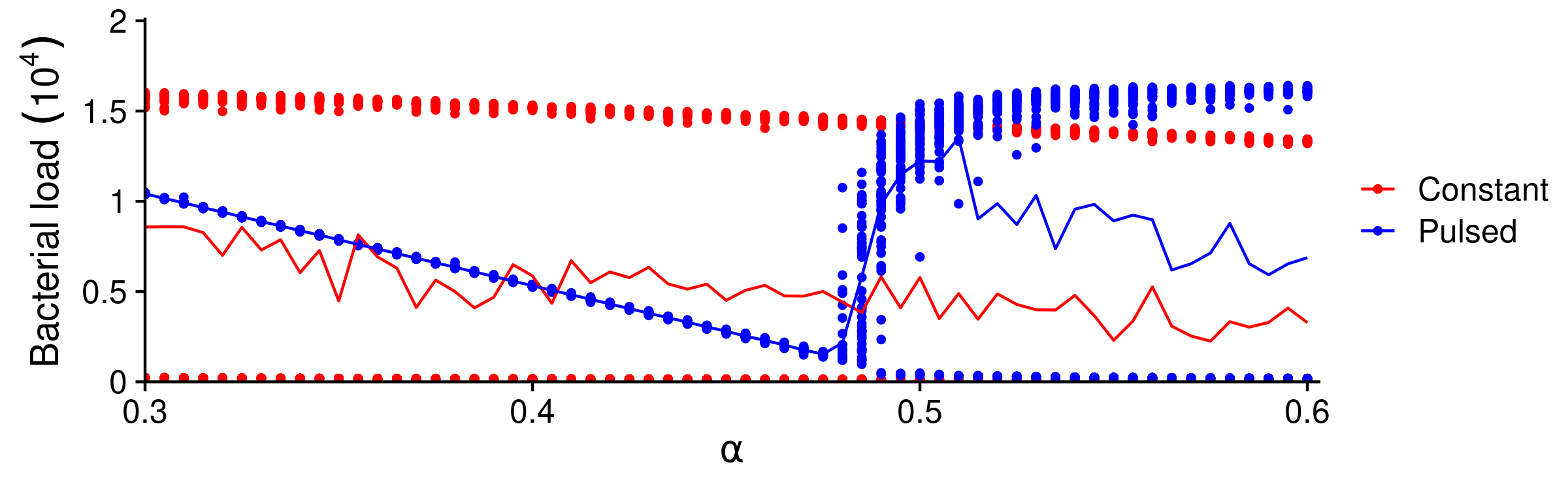}
        \end{subfigure}
        \caption{The average bacterial load over time for constant and pulsed ($2$hrs on and off each) for various $\alpha$. The points are the results for individual realizations and the curves their average. Increasing $\kappa$ from $\kappa=4$ (panel \textbf{a}) to $\kappa=20$ (panel \textbf{b}) shifts the minimum of the average bacterial load to the right.}
        \label{fig:alpha_k20}
    \end{figure}

Figure \ref{fig:alpha_k20} shows how increasing the competitiveness $\kappa$ can shift the minimum of the average bacterial load to the right. Thus, we can counteract the effect of a high antibiotic kill rate. Thereby, competition from the wild-type can be used to effectively suppress the resistant mutants.

\begin{figure}[!htbp]
    \centering
    \begin{subfigure}[]{0.3\textwidth}
        \caption{}
        \includegraphics[width=\textwidth]{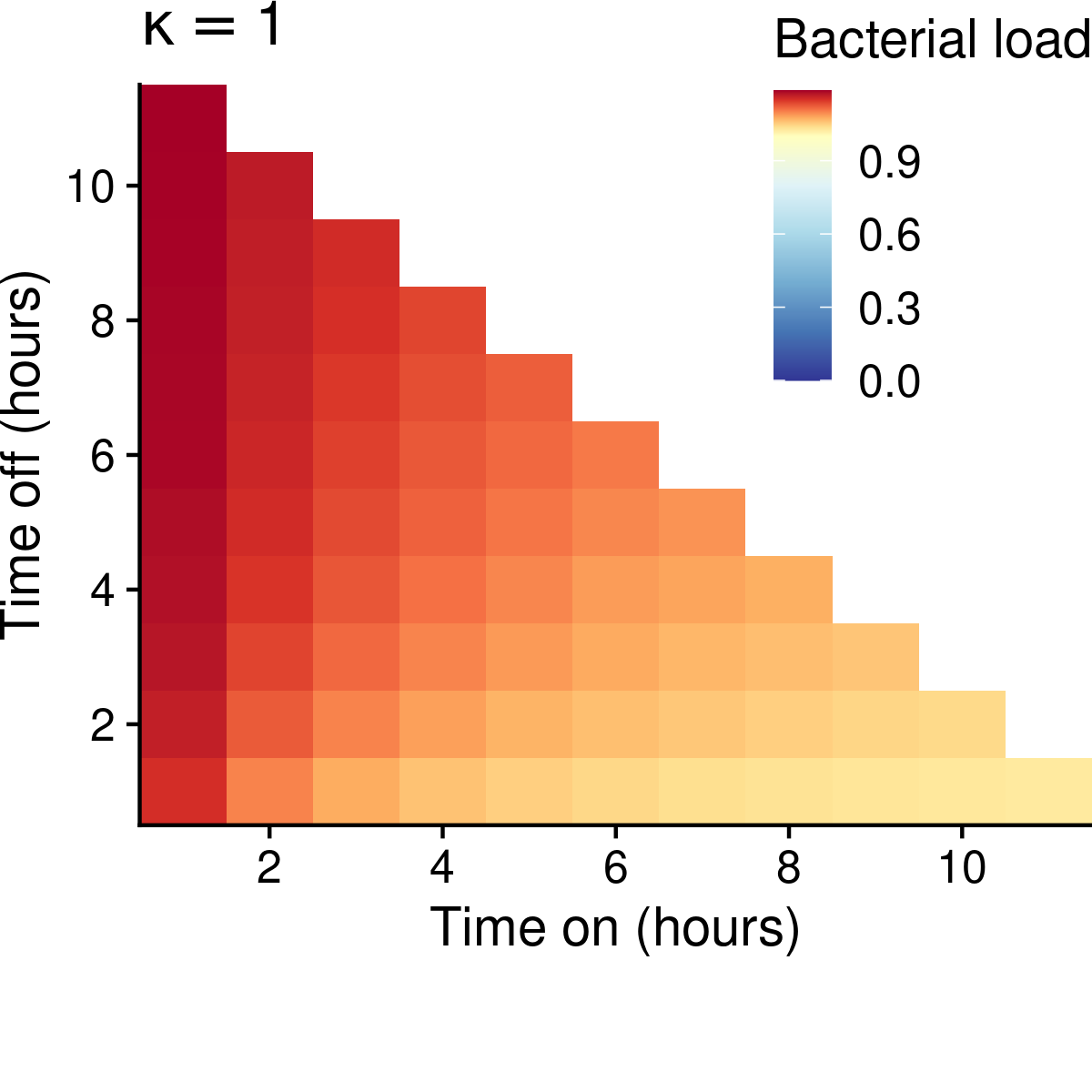}
    \end{subfigure}
    \begin{subfigure}[]{0.3\textwidth}
        \caption{}
        \includegraphics[width=\textwidth]{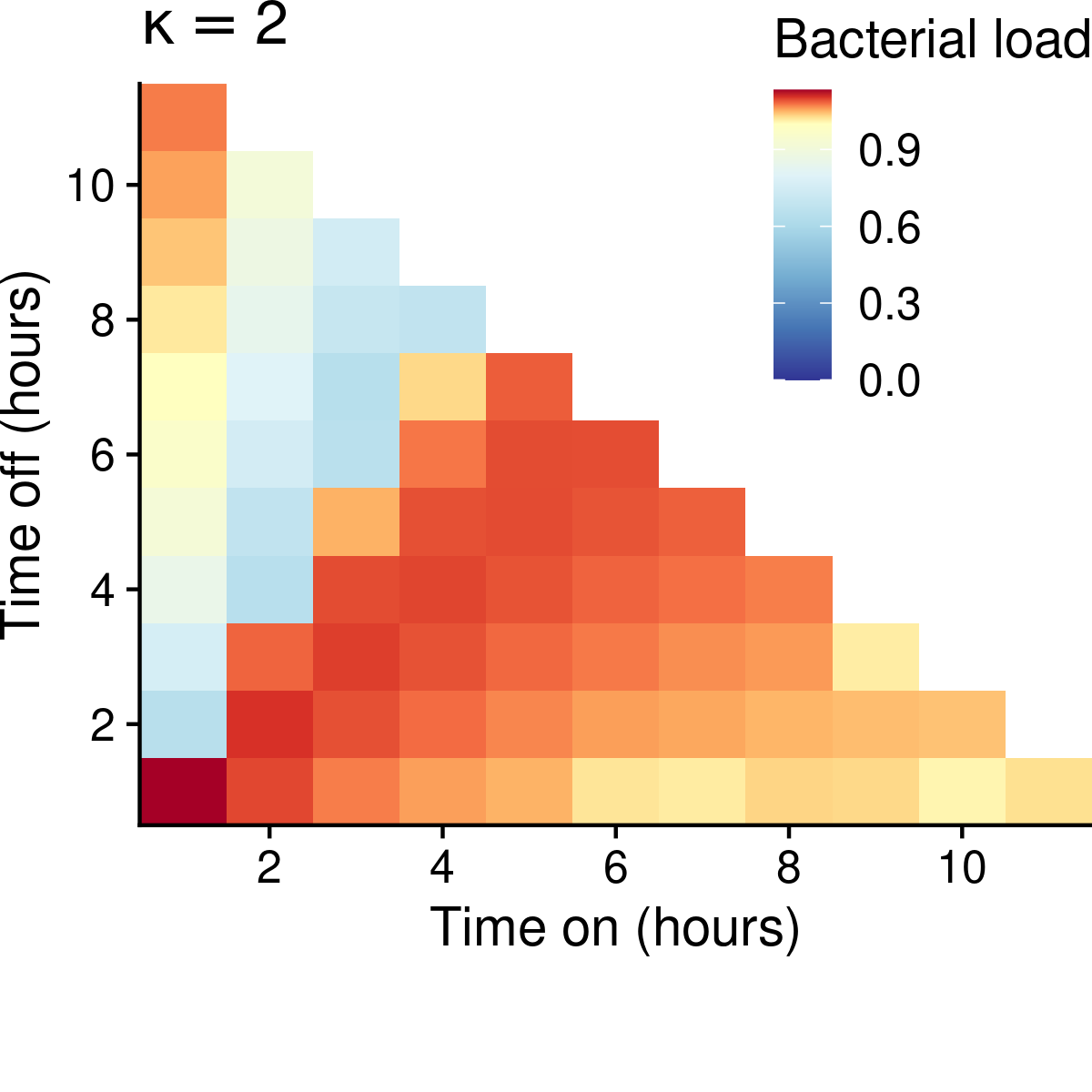}
    \end{subfigure}
    \begin{subfigure}[]{0.3\textwidth}
        \caption{}
        \includegraphics[width=\textwidth]{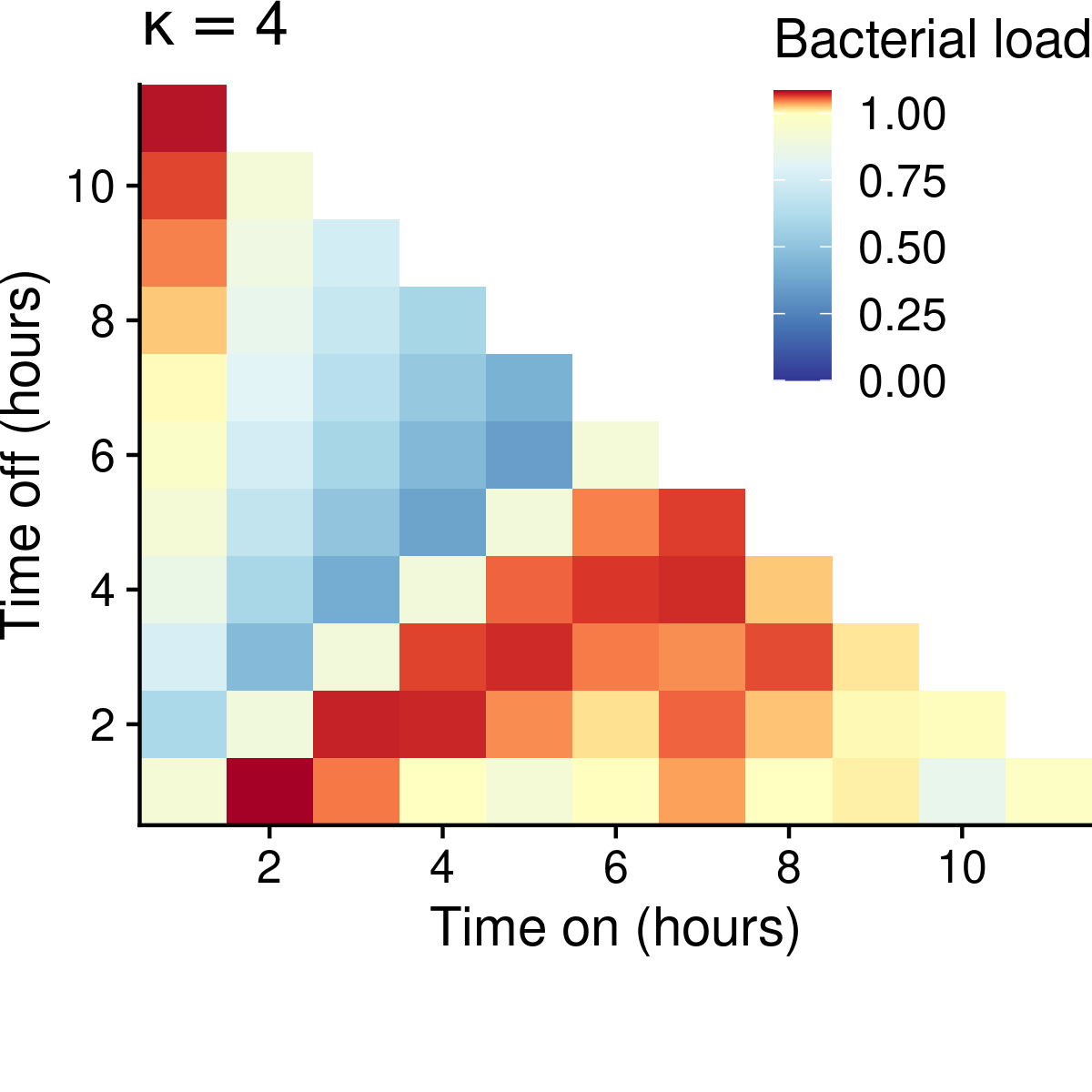}
    \end{subfigure} \\
    \begin{subfigure}[]{0.3\textwidth}
        \caption{}
        \includegraphics[width=\textwidth]{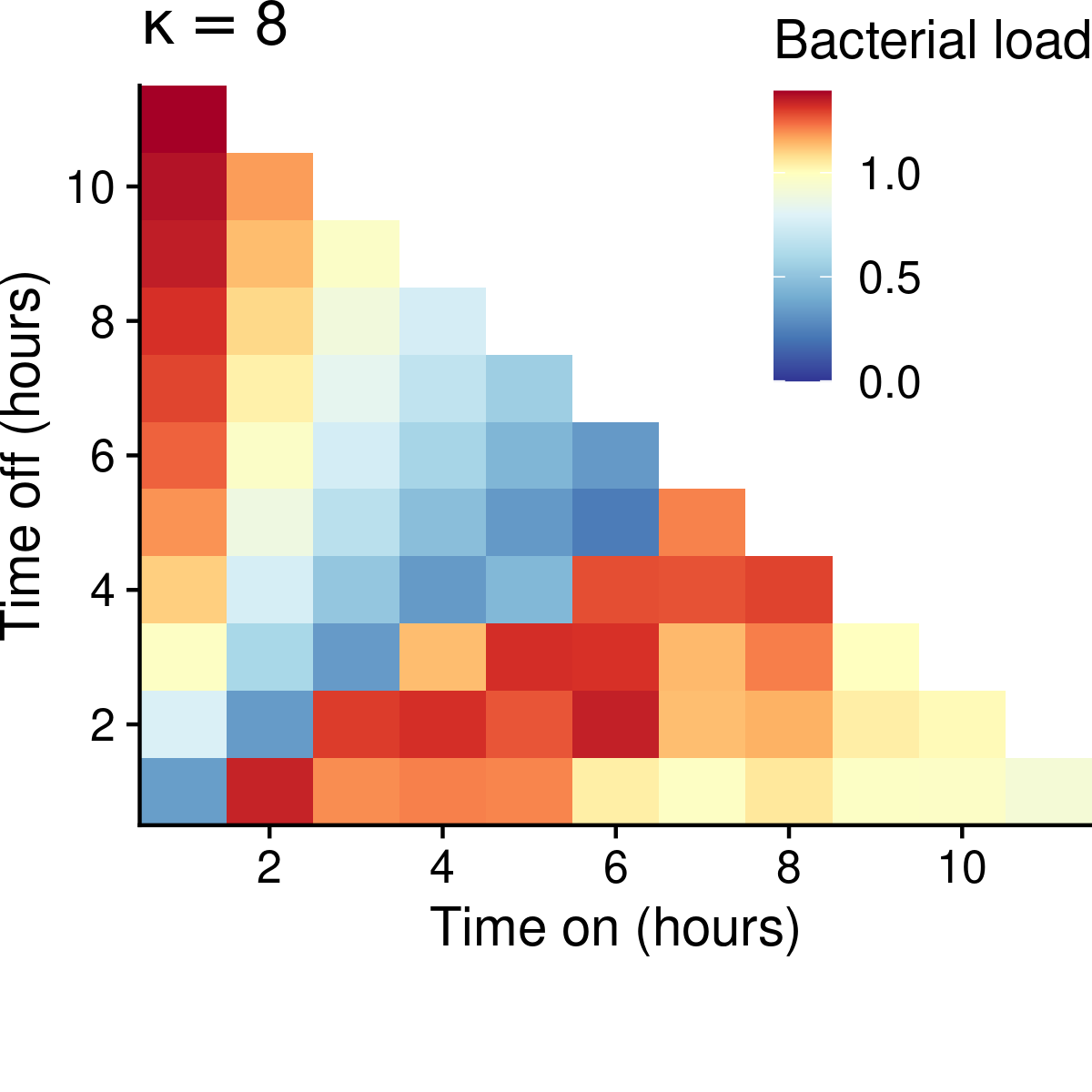}
    \end{subfigure}
    \begin{subfigure}[]{0.3\textwidth}
        \caption{}
        \includegraphics[width=\textwidth]{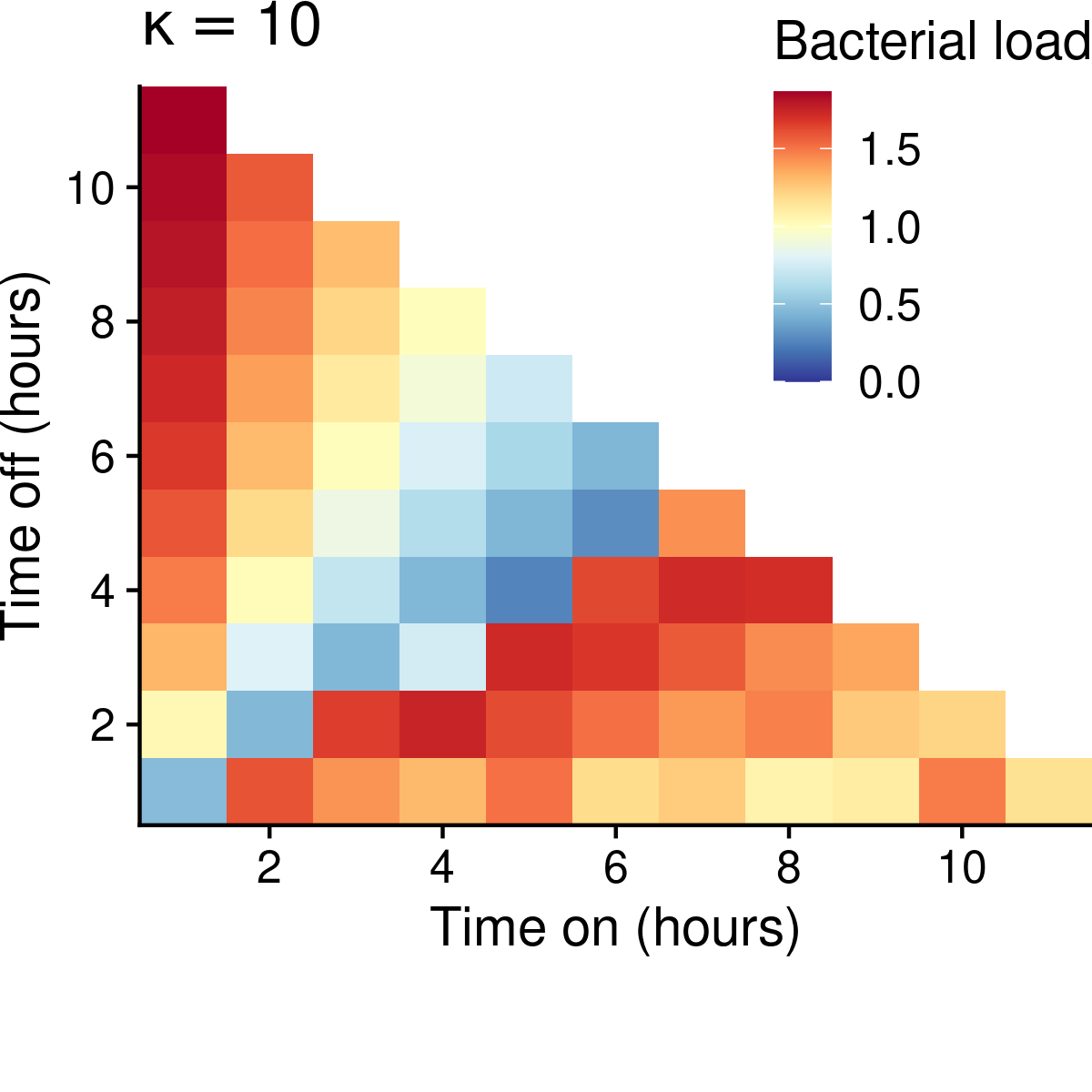}
    \end{subfigure}
    \begin{subfigure}[]{0.3\textwidth}
        \caption{}
        \includegraphics[width=\textwidth]{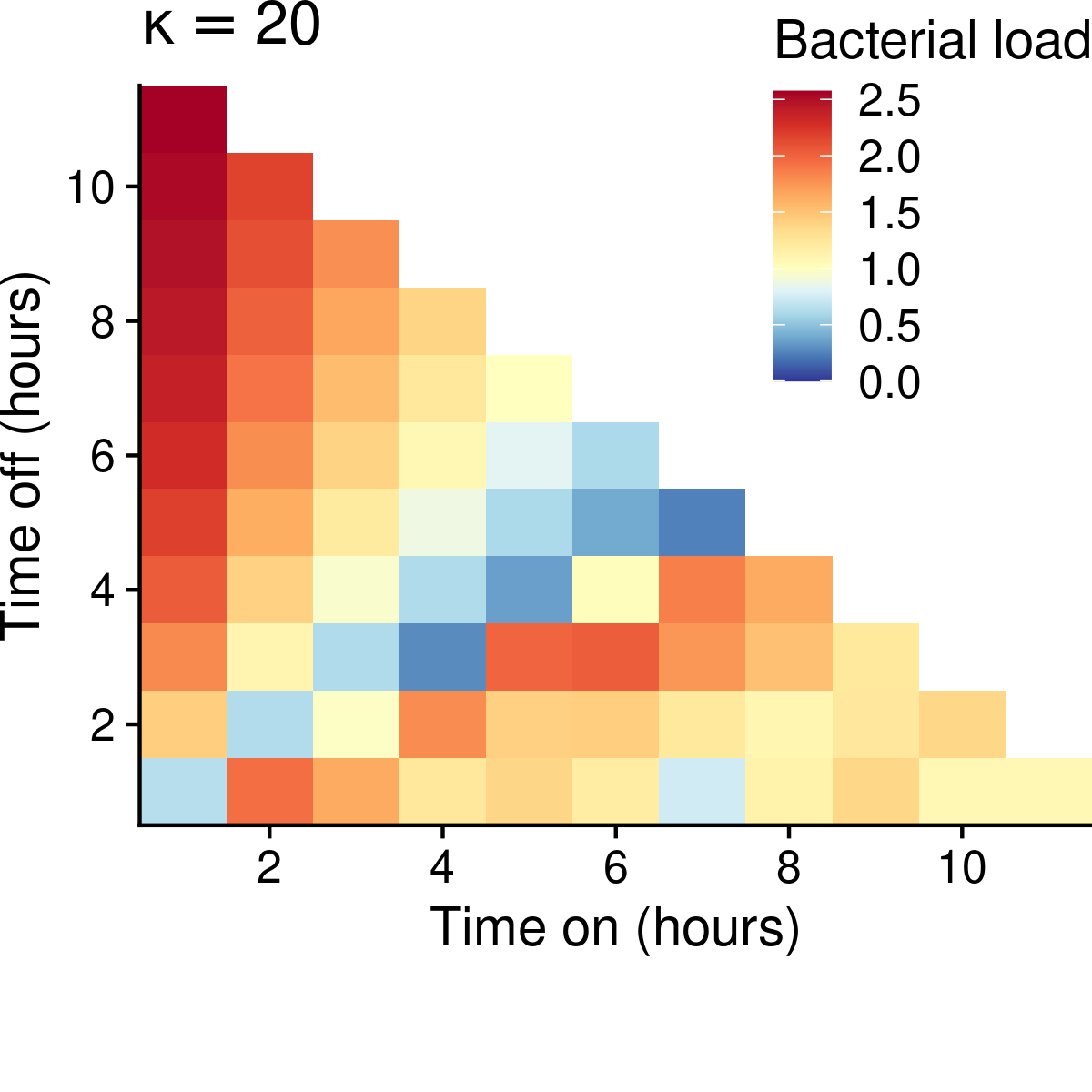}
    \end{subfigure} \\
    \caption{Heatmaps of the average bacterial load over time from pulsed protocols relative to that of constant application of the antibiotic for $\kappa= 1, 2, 4, 8, 10, 20$. Protocols that matched the average outcome of the constant application therapy are coloured in yellow. Those protocols that did worse are in red, and those that did better are in blue. Here the cost to resistance is zero ($c=0$).}
    \label{fig:zerocost}
\end{figure}

Though compensatory mutations could remove the cost to resistance and thus reduce the effectiveness of pulsed protocols, if there is a competitive disadvantage, resistance can still be mitigated. This results can be seen in Figure \ref{fig:zerocost}, which depicts the results for various $\kappa$ and zero cost of resistance.

\end{document}